\documentstyle[12pt,aaspp4]{article}
\def\tempest%
{\begin{array}{ccc}
1 & 1 & 1 \\
1 & 1 & 1 \\
4 & 3 & 8
\end{array}}

\def\kms{{\rm km}\,{\rm s}^{-1}}
\def\kpc{{\rm kpc}}
\def\day{{\rm day}}

\def\lim{{\rm lim}}
\def\dls{{D_{\rm LS}}}
\def\dos{{D_{\rm S}}}
\def\dol{{D_{\rm L}}}

\def\cc{{\rm cc}}

\begin{document}

\title{Combined Analysis of the Binary-Lens Caustic-Crossing Event
MACHO 98-SMC-1}

\author{
C.~Afonso\altaffilmark{1}, 
C.~Alard\altaffilmark{2},
J.N.~Albert\altaffilmark{3},
J.~Andersen\altaffilmark{4},
R.~Ansari\altaffilmark{3}, 
\'E.~Aubourg\altaffilmark{1}, 
P.~Bareyre\altaffilmark{1,5}, 
F.~Bauer\altaffilmark{1},
J.P.~Beaulieu\altaffilmark{6,7,8},
A.~Bouquet\altaffilmark{5},
S.~Char\altaffilmark{9},
X.~Charlot\altaffilmark{1},
F.~Couchot\altaffilmark{3}, 
C.~Coutures\altaffilmark{1}, 
F.~Derue\altaffilmark{3}, 
R.~Ferlet\altaffilmark{6},
J.F.~Glicenstein\altaffilmark{1},
B.~Goldman\altaffilmark{1,10,11},
A.~Gould\altaffilmark{1,8,12},
D.~Graff\altaffilmark{1,12},
M.~Gros\altaffilmark{1}, 
J.~Haissinski\altaffilmark{3}, 
J.C.~Hamilton\altaffilmark{5},
D.~Hardin\altaffilmark{1},
J.~de Kat\altaffilmark{1}, 
A.~Kim\altaffilmark{5},
T.~Lasserre\altaffilmark{1},
\'E.~Lesquoy\altaffilmark{1},
C.~Loup\altaffilmark{6},
C.~Magneville \altaffilmark{1}, 
J.B.~Marquette\altaffilmark{6},
\'E.~Maurice\altaffilmark{13}, 
A.~Milsztajn \altaffilmark{1},  
M.~Moniez\altaffilmark{3},
N.~Palanque-Delabrouille\altaffilmark{1}, 
O.~Perdereau\altaffilmark{3},
L.~Pr\'evot\altaffilmark{13}, 
N.~Regnault\altaffilmark{3},
J.~Rich\altaffilmark{1}, 
M.~Spiro\altaffilmark{1},
A.~Vidal-Madjar\altaffilmark{6},
L.~Vigroux\altaffilmark{1},
and
S.~Zylberajch\altaffilmark{1}
}
\author{(The EROS Collaboration)}
\author{ 
  C.~Alcock\altaffilmark{14,15},           
  R.A.~Allsman\altaffilmark{16},          
  D.~Alves\altaffilmark{14,17,37},            
  T.S.~Axelrod\altaffilmark{18},          
  A.C.~Becker\altaffilmark{15,19,20},         
  K.H.~Cook\altaffilmark{14},             
  A.J.~Drake\altaffilmark{18},            
  K.C.~Freeman\altaffilmark{18},          
  K.~Griest\altaffilmark{15,22},           
  L.J.~King\altaffilmark{15,21,23},           
  M.J.~Lehner\altaffilmark{24},           
  S.L.~Marshall\altaffilmark{14},         
  D.~Minniti\altaffilmark{14,25},         
  B.A.~Peterson\altaffilmark{18},         
  M.R.~Pratt\altaffilmark{26},           
  P.J.~Quinn\altaffilmark{27},           
  A.W.~Rodgers\altaffilmark{18},          
  P.B.~Stetson\altaffilmark{28},         
  C.W.~Stubbs\altaffilmark{15,19},         
  W.~Sutherland\altaffilmark{29},        
  A.~Tomaney\altaffilmark{19},            
and
  T.~Vandehei\altaffilmark{15,22}         
}
\author{(The MACHO/GMAN Collaboration)}
\author{
  S.H.~Rhie\altaffilmark{21,30},             
  D.P.~Bennett\altaffilmark{14,15,21,30},    
  P.C.~Fragile\altaffilmark{21},             
  B.R.~Johnson\altaffilmark{31},             
and
   J.Quinn\altaffilmark{21,30}               
}
\author{(The MPS Collaboration)}
\author{
A.~Udalski\altaffilmark{32},
M.~Kubiak\altaffilmark{32},
M.~Szyma{\'n}ski\altaffilmark{32},\\ 
G.~Pietrzy\'nski\altaffilmark{32}, 
P.~Wo\'zniak\altaffilmark{33}, 
and
K.~\.Zebru\'n\altaffilmark{32}
}
\author{(The OGLE Collaboration)}
\author{
M.D.~Albrow\altaffilmark{34},
J.A.R.~Caldwell\altaffilmark{35}, 
D.L.~DePoy\altaffilmark{12}, 
M.~Dominik\altaffilmark{7}, 
B.S.~Gaudi\altaffilmark{12}, 
J.~Greenhill\altaffilmark{36}, 
K.~Hill\altaffilmark{36},
S.~Kane\altaffilmark{36,37}, 
R.~Martin\altaffilmark{38},  
J.~Menzies\altaffilmark{35}, 
R.M.~Naber\altaffilmark{7}, 
R.W.~Pogge\altaffilmark{12},
K.R.~Pollard\altaffilmark{34},
P.D.~Sackett\altaffilmark{7}, 
K.C.~Sahu\altaffilmark{37}, 
P.~Vermaak\altaffilmark{35},  
R.~Watson\altaffilmark{36}, 
and
A.~Williams\altaffilmark{38}
}
\author{(The PLANET Collaboration)}

\altaffiltext{1}{
CEA, DSM, DAPNIA,
Centre d'\'Etudes de Saclay, 91191 Gif-sur-Yvette Cedex, France}
\altaffiltext{2}{
DASGAL, 77 avenue de l'Observatoire, 75014 Paris, France}
\altaffiltext{3}{
Laboratoire de l'Acc\'{e}l\'{e}rateur Lin\'{e}aire,
IN2P3 CNRS, Universit\'e Paris-Sud, 91405 Orsay Cedex, France}
\altaffiltext{4}{
Astronomical Observatory, Copenhagen University, Juliane Maries Vej 30, 
2100 Copenhagen, Denmark}
\altaffiltext{5}{
Coll\`ege de France, Physique Corpusculaire et Cosmologie, IN2P3 CNRS, 
11 pl.~Marcellin Berthelot, 75231 Paris Cedex, France}
\altaffiltext{6}{
Institut d'Astrophysique de Paris, INSU CNRS,
98bis Boulevard Arago, 75014 Paris, France}
\altaffiltext{7}{Kapteyn Astronomical Institute, Postbus 800, 
9700 AV Groningen, The Netherlands}
\altaffiltext{8}{
PLANET Collaboration member}
\altaffiltext{9}{
Universidad de la Serena, Facultad de Ciencias, Departamento de Fisica,
Casilla 554, La Serena, Chile}
\altaffiltext{10}{
Dept.~Astronom\'\i a, Universidad de Chile, Casilla 36-D, Santiago,
Chile}
\altaffiltext{11}{
European Southern Observatory, Casilla 19001, Santiago 19, Chile}
\altaffiltext{12}{
Department of Astronomy, Ohio State University, Columbus, OH 43210, USA.}
\altaffiltext{13}{
Observatoire de Marseille,
2 pl.~Le Verrier, 13248 Marseille Cedex 04, France}
\altaffiltext{14}{Lawrence Livermore National Laboratory, Livermore, CA 94550}
\altaffiltext{15}{Center for Particle Astrophysics,
  University of California, Berkeley, CA 94720}
\altaffiltext{16}{Supercomputing Facility, Australian National University,
  Canberra, ACT 0200, Australia }
\altaffiltext{17}{Department of Physics, University of California, Davis, CA 95616}
\altaffiltext{18}{Mt.Stromlo and Siding Spring Observatories,
  Australian National University, Weston, ACT 2611, Australia}
\altaffiltext{19}{Departments of Astronomy and Physics,
  University of Washington, Seattle, WA 98195}
\altaffiltext{20}{MPS Collaboration member}
\altaffiltext{21}{Department of Physics, University of Notre Dame, Notre Dame, IN 46556}
\altaffiltext{22}{Department of Physics, University of California,
  San Diego, CA 92093}
\altaffiltext{23}{Max-Planck-Institut f\"ur Astrophysik, Karl Schwarzschild Str 1, Postfach 1523, D-85740 Garching, Germany}
\altaffiltext{24}{Department of Physics, University of Sheffield,
Sheffield s3 7RH, UK}
\altaffiltext{25}{Departmento de Astronomia, P.~Universidad Cat\'olica, 
Casilla 104, Santiago 22, Chile} 
\altaffiltext{26}{Center for Space Research, MIT,
Cambridge, MA 02139}
\altaffiltext{27}{European Southern Observatory, Karl-Schwarzchild Str. 2, D-857
48, Garching, Germany}
\altaffiltext{28}{National Research Council,
  5071 West Saanich Road, RR 5, Victoria, BC V8X 4M6, Canada}
\altaffiltext{29}{Department of Physics, University of Oxford,
  Oxford OX1 3RH, U.K.}
\altaffiltext{30}{MACHO/GMAN Collaboration member}
\altaffiltext{31}{Tate Laboratory of Physics, University of Minnesota, 
  Minneapolis, MN 55455}
\altaffiltext{32}{Warsaw University Observatory, Al.~Ujazdowskie~4, 00-478~Warszawa, Poland}
\altaffiltext{33}{Princeton University Observatory, Princeton, NJ 08544-1001, USA}
\altaffiltext{34}{Univ. of Canterbury, Dept. of Physics \& Astronomy, 
Private Bag 4800, Christchurch, New Zealand}
\altaffiltext{35}{South African Astronomical Observatory, P.O. Box 9, 
Observatory 7935, South Africa}
\altaffiltext{36}{Univ. of Tasmania, Physics Dept., G.P.O. 252C, 
Hobart, Tasmania~~7001, Australia}
\altaffiltext{37}{Space Telescope Science Institute, 3700 San Martin Drive, 
Baltimore, MD 21218~~U.S.A.}
\altaffiltext{38}{Perth Observatory, Walnut Road, Bickley, Perth~~6076, Australia}


\begin{abstract}

	We fit the data for the binary-lens microlensing event
MACHO 98-SMC-1 from 5 different microlensing collaborations and find
two distinct solutions characterized by binary separation $d$ and mass
ratio $q$:  $(d,q)=(0.54,0.50)$ and $(d,q)=(3.65,0.36)$, where 
$d$ is in units of the Einstein radius.
However, the relative proper motion of the lens is very similar in the two
solutions, $1.30\,\kms\,\kpc^{-1}$ and $1.48\,\kms\,\kpc^{-1}$,
thus confirming that the lens is in the Small Magellanic Cloud.
The close binary can be either rotating or approximately
static but the wide binary must be rotating at close its maximum allowed rate
to be consistent with all the data.
We measure limb-darkening coefficients for five bands ranging from 
$I$ to $V$.  As expected, these progressively decrease with rising wavelength.
This is the first measurement of limb darkening for a metal-poor A star.

\end{abstract}


\newpage
\section{Introduction}

	The binary-lens microlensing event MACHO 98-SMC-1, found by the MACHO 
collaboration (Alcock et al.\ 1999a) in observations toward the Small 
Magellanic Cloud (SMC), was observed by five different groups.   Each
group attempted to measure or constrain the relative lens-source proper
motion by fitting the observed light curve to a binary-lens model, and
each concluded
that the proper motion is consistent with the lens being in the SMC rather
than the Galactic halo (Afonso et al.\ 1998 [EROS]; 
Albrow et al.\ 1999a [PLANET]; Alcock et al.\ 1999a [MACHO/GMAN]; 
Udalski et al.\ 1998b [OGLE]; Rhie et al.\ 1999 [MPS]).  
Despite this unanimous opinion, there are several reasons for taking a closer 
look at this event.  

	First, Albrow et al.\ (1999c) have subsequently developed a more
robust method for finding binary-lens solutions that are consistent with a 
given light-curve data set.  They found a broad set of degeneracies for the
fit to MACHO 98-SMC-1 based on PLANET data and showed that based on these data
alone, the proper motion could not be constrained to better than a factor of
four.  Such degeneracies are likely to be endemic to binary-lens fitting
(Dominik 1999a).
Albrow et al.\ (1999c) showed that by including additional data it would
be possible to remove at least some of these degeneracies, but they argued
that one major ambiguity (between ``wide'' and ``close'' binaries) might be
difficult to resolve.  Hence, it is important to determine whether these
degeneracies can in fact be resolved by combining all available data.
In particular, the OGLE and MPS data together constrain the first caustic
crossing, the MACHO data constrain the baseline, the PLANET data give 
excellent coverage of the main part of the second caustic crossing, and
the EROS data give similar coverage of the end of the second caustic crossing.

	Second, caustic-crossing binary events allow one in principle to
measure the limb-darkening profile of the source star, as Albrow et al.\ 
(1999b) have done for a K giant using MACHO 97-BLG-28.
The source in MACHO 98-SMC-1 is an A star as determined both from its colors
(Albrow et al.\ 1999a; Rhie et al.\ 1999) and its spectrum
(Albrow et al.\ 1999a).  Since it is in
the SMC, it is almost certainly metal poor.  If the limb darkening were 
measured, it would be the first such measurement for a metal-poor A star.
The caustic crossing occurred while the source was visible from South Africa.
The PLANET SAAO data have excellent coverage of the main part of the
caustic crossing, and it 
therefore might be possible to measure the limb darkening from these 
data alone using a variant of the method of Albrow et al.\ 
(1999c) which is described more fully in \S\ 3.  However, it is not clear
to what degree the degeneracies in the overall fit 
would compromise such a determination.  By fitting all the data
these degeneracies could be partially or totally removed.  
The EROS data from Chile provide excellent coverage of the end of the 
crossing and the MACHO/GMAN data also cover the end of the crossing.  
Neither of these data sets can determine
the limb darkening without fixing the characteristics of the caustic crossing,
which in turn requires the PLANET data.

	Third, the light-curve coverage of MACHO 98-SMC-1 is one of the best 
of any binary-lens event yet observed.  It is therefore an excellent laboratory
in which 
to search for additional, unanticipated anomalies that may be present in
microlensing events but have not yet been noticed.

	In \S\ 2 we briefly review the data that are available for this
event.  In \S\ 3 we summarize and extend the Albrow et al.\ (1999c) method
for finding binary-lens
solutions.  In \S\ 4 we present our results for static binaries, 
including measurement of the limb-darkening coefficients.  In \S\ 5, we
derive the proper motion which in turn determines the projected separation
of the binary.  We use the latter quantity to constrain the period of
binary orbit.  We consider rotating binary models that satisfy this
constraint in \S\ 6.  Finally, in \S\ 7, we study the relationship of the
solutions derived here to those reported in earlier investigations that
were based on subsets of our combined data set.  We show that all the 
close-binary solutions are in fact different positions within one broad minimum
in $\chi^2$.  The wide-binary solutions of Albrow (1999c) represent another
broad minimum.  We also resolve some puzzling discrepancies between different
solutions.

\section{Data}

	We describe all dates using HJD$'= $HJD$- 2450000.0$ where HJD is the
Heliocentric Julian Date.  The reported times are the midpoints of the
exposures. We combine a total of 14 light curves obtained at 8 different
telescopes.  These were reduced using various photometry packages as described
below.  In all cases, points that failed internal tests of these packages
were eliminated prior to beginning the fitting process.

	The first two light curves were taken in 
the (broad non-standard) MACHO $R$ and MACHO $B$
filters on the 1.3 m telescope at the Mount Stromlo \& Siding Springs 
Observatory (MSSSO) near Canberra, Australia.  These contain 727 and and 735
points respectively, beginning about 5 years before the event and ending
 81 days after the second caustic crossing at HJD$'\sim 982.6$.  
The Mt.\ Stromlo 1.3 m is normally used
to search for microlensing events and hence typically takes one exposure
per clear night.  However, because of the importance of this event, a total 
of 23
exposures were obtained during the nights just before and after the second
caustic crossing.  The next two curves were taken in the
(standard Johnson/Cousins) $R$
and $B$ filters on the 0.9 m telescope at the Cerro Tololo Interamerican
Observatory (CTIO) near La Serena, Chile.  These contain 83 and 22 points
respectively, beginning 7 days before the first caustic crossing
at HJD$'\sim 970.5$ and ending 6 days after the second crossing.
MACHO has 1 hour per night on this telescope which was mostly dedicated to
MACHO 98-SMC-1 during this period of observation.  All four of these
light curves were reported by Alcock et al.\ (1999a).  However, all the
points after the first caustic crossing of the MSSSO data
were re-reduced using image subtraction (Tomaney 1998)
which we determined has somewhat 
smaller errors than the original SoDoPhot reductions.  These include 
respectively 16 and 21 late-time points (more than 24 days after the second
caustic crossing) that were not previously reported by Alcock et al.\ (1999a).

	Next is the OGLE (standard Cousins) I band curve from the 1.3 m
Warsaw telescope at Las Campanas,
Chile.  The images are from OGLE's routine monitoring of the SMC to search
for microlensing events, and OGLE made no special effort to observe this
event.  Rather, they analyzed their images after the event was over and
found 7 measurements during the interval from 4 days before the first
caustic crossing to 4 days after the second.  As discussed by Udalski et al.\
(1998b), the primary interest in this relatively small data set comes from
the second data point on HJD$'=970.9037$ which is highly magnified 
($A\sim 29$) and therefore comes either just before or just after the first 
caustic crossing.  

	The sixth and seventh curves were taken in 
(broad non-standard) EROS $R$ and EROS $B$
filters on the 1.0 m Marly telescope at the European Southern Observatory
at La Silla, Chile.  Normally, this telescope is operated in survey mode to
search for microlensing events.  However, it was down for maintenance during 
most of the time that MACHO 98-SMC-1 was inside the caustic, and 
recommenced operations only on the night of the second caustic crossing.
In light of the importance of MACHO 98-SMC-1, the telescope was entirely
dedicated to observing this event 
during this night and made several observations
per night for the next 15 nights, whereupon it resumed normal operations.
The observations of the first night were previously reported by
Afonso et al.\ (1998).  The rest of the observations are reported here
for the first time.  All the observations have been re-reduced using
an improved version (Alard 1999) of the algorithm used 
by Afonso et al.\ (1998).
There were a total of 111 observations
in $R$ and 131 in $B$.  These include about 8 points in each band 
from two years before
the event and about another 8 from the year after the event when
the source is approaching baseline.

	The eighth curve is based on the (standard Cousins) $R$ band 
observations taken by the MPS collaboration using 
the 1.9 m telescope at MSSSO.  A total of 34 observations were taken from
just before the first caustic crossing until 4 days after the second.
In addition there is one late-time baseline measurement taken 67 days after
the caustic crossing.  These observations were earlier reported by
Rhie et al.\ (1999).  Of particular note is the first observation on
HJD$'=970.0485$ which is clearly before the caustic crossing.  This data
point, combined with the OGLE data point 0.9 days later, strongly constrain
the time of the first crossing.

	The ninth through thirteenth curves are based on standard Cousins $I$
band observations taken by the PLANET collaboration using the 
SAAO 1 m telescope at Sutherland, 
South Africa, the Yale-CTIO 1 m, the CTIO 0.9 m,
and the Canopus 1 m near Hobart, Tasmania.  The SAAO data are divided into
two groups because of a change in the CCD detector
at HJD$'=980.0$, two days before the
caustic crossing.  The five PLANET data sets
comprise respectively
13, 175, 32, 13, and 1 observations.  The CTIO 0.9 m data cover only the 
interior of the caustic, beginning 3 days after the first crossing and ending
4 days before the start of the second.  The Canopus data contain only 1 point
about 1 day before the second crossing.  The Yale-CTIO 1 m data
begin 4 days before the second crossing and end 14 days after it.  The
SAAO 1 m data begin 5 days after the first caustic and end 44 days after the
second.  Of all the observations, only the SAAO data cover the peak of the
second caustic crossing.  Moreover, they do so quite densely.  Most of these
data were previously reported by Albrow et al.\ (1999a, 1999c).  
However, all of the SAAO data after HJD$'=980.0$ have been re-reduced using
image subtraction (Alard 1999)
which we found produces significantly lower errors and
fails significantly less frequently than even the best DoPhot reductions
previously reported by Albrow et al.\ (1999c).  Details of this comparison
will be given elsewhere.  In addition, we have eliminated
the SAAO data from the first night, HJD$'=975$, because there was yet another
CCD change on  HJD$'=976.0$ rendering the conditions on this night unique.

	Finally, the fourteenth light curve is based on standard Johnson $V$
band observations taken by the PLANET collaboration using the SAAO 1 m.
These comprise 24 observations including 14 taken during the second caustic
crossing and 10 taken over the next 32 days.  These data were used by
Albrow et al.\ (1999a) to determine the $V$ brightness of the source and so
its $V-I$ color, but have not previously been made available.

	In a preliminary fit to the data, we find that four data points
are significant outliers.  These are the SAAO point at 
HJD$'=979.6424$, the MACHO $B$ point at
HJD$'=982.2061$
the EROS $R$ point at
HJD$'=982.8427$,
and  the MACHO $B$ point at
HJD$'=997.1607$,
with residuals of $-5.4$, $-4.9$, $-4.1$, and $3.9\,\sigma$, respectively
We eliminate these from future modeling.  We renormalize the quoted errors from
each light curve by a factor so as to force $\chi^2/$dof (degree of freedom)
to be unity for
that curve.  The factors are
MACHO $R$ (SoDoPhot): 1.12,
MACHO $B$ (SoDoPhot): 1.12,
MACHO $R$ (image subtraction): 1.26,
MACHO $B$ (image subtraction): 1.58,
MACHO-CTIO  $R$: 0.94,
MACHO-CTIO  $B$: 1.10,
OGLE  $I$: 1.00,
EROS  $R$: 1.32,
EROS  $B$: 0.96,
MPS   $R$: 1.80,
PLANET-SAAO (HJD$'<980)$  $I$: 1.04,
PLANET-SAAO (HJD$'>980)$  $I$: 0.97,
PLANET-Yale-CTIO          $I$: 0.97,
PLANET-CTIO               $I$: 0.90, and
PLANET-SAAO               $V$: 2.21.
The PLANET-Canopus $I$ was not renormalized because there is only one point.
The precise value of the renormalization factors depends slightly on which
solution is adopted, but we find that our results are not sensitive
to these small changes.  We bin the early MACHO $R$ and $B$ data in 20-day
intervals (see Fig.\ 1, below).  With this binning, there are a total of
1018 data points.

\section{Method}

	To analyze these data, we follow and slightly extend the method
of Albrow et al.\ (1999c).  We first review this method and its
motivation and then discuss its extension.

	Events where a non-rotating binary lens passes
in front of a uniform finite source are described by
$(7+2n)$ parameters, where $n$ is the number of observatories: 
three parameters correspond to the three geometrical parameters of a 
point-like single
lens (Einstein time scale $t_{\rm E}$, impact parameter $u_0$, 
and time of closest approach $t_0$), 
three other parameters 
characterize its binary nature (mass ratio $q$, separation $d$ in units of the
Einstein radius, and angle $\alpha$ of the source trajectory relative to the 
binary axis), one, $\rho_*$, describes the
size of the source relative to the Einstein ring, and there
are $n$ parameters 
for the source flux, $F_{\rm s,1} \ldots F_{{\rm s},n}$,
and $n$ for the unlensed background light $F_{\rm b,1} \ldots F_{{\rm b},n}$,
that is one pair for each of the $n$ observatories.  For events where the
source crosses a fold caustic, one may define several additonal useful
parameters which can be derived from these $7+2n$, including 
the position within the Einstein ring where the source
center crosses the caustic, ${\bf u}_\cc$, the time of the caustic crossing,
$t_\cc$, the angle $\phi$
of the source trajectory relative to caustic at the crossing, 
the half-duration of the crossing $\Delta t$, and the radius crossing time
$t_*\equiv \Delta t\sin\phi$.
Generally, it is not difficult to find a set of parameters
that yield a satisfactory fit to the data.  However, it is often unclear
whether there exist other equally good or better fits.  One would like
to make a systematic search through parameter space but because of the
size and complexity of the parameter space, a brute-force search is out
of the question.

	Albrow et al.\ (1999c) showed that for events with a
well-observed caustic crossing, it is possible to greatly reduce the
space of allowed solutions thereby permitting a systematic search of
the remaining parameter space.  The method proceeds in three steps.
First, the caustic crossing is fit to a 5-parameter function.  Second,
these parameters are used to constrain a coarse-grained but systematic
search through parameter space for solutions that can accomodate the
non-caustic-crossing data.  Third, final solution(s) are found by
$\chi^2$ minimization using the results from the coarse-grained search
as initial guesses.

	In the first step, the
light curve is fit to a 5-parameter curve of the form
\begin{equation}
F(t) = \biggl({Q\over\Delta t}\biggr)^{1/2} 
G_0\biggl({t - t_{\cc}\over\Delta t}\biggr)
+ F_\cc + \tilde \omega(t - t_{\cc}),\label{eqn:fform}
\end{equation}
where
\begin{equation}
G_0(\eta) \equiv 
\frac{2}{\pi}\int_{\rm max(\eta,-1)}^{1}d x\,\left(\frac{1-x^2}{x-\eta}
\right)^{1/2}\, \Theta(1-\eta),\label{eqn:gdef}
\end{equation}
is the normalized light curve of a (second) 
caustic crossing with a uniform source,
and $\Theta$ is a step function.
Here 
$Q$ is related to the rise
time of the caustic (defined more precisely below), $F_\cc$ is the magnified
flux from the source when it is immediately outside the caustic, and
$\tilde \omega$ is the slope of the light curve immediately outside the
caustic.
Using the PLANET data, Albrow et al.\ (1999c) found
\begin{equation}
Q= (15.73\pm 0.35)\, F_{20}^2\,\day,\quad
t_{\cc} = (982.62439\pm 0.00087)\,\day,\quad
\Delta t = (0.1760\pm 0.0015)\,\day,
\label{eqn:parmfit1}
\end{equation}
\begin{equation}
F_\cc = (1.378\pm 0.096)\,F_{20},\qquad
\tilde \omega = (0.02\pm 0.10)\,F_{20}\,\day^{-1},
\label{eqn:parmfit2}
\end{equation}
where $F_{20}$ is the flux from an $I=20$ star.  For the first step, we
simply adopt the results summarized in equations (\ref{eqn:parmfit1})
and (\ref{eqn:parmfit2}).

	In the second step, the search of the full parameter space is
substantially narrowed by making use of these caustic-crossing parameters
with the following relations between observed and theoretical quantities,
\begin{equation}
F_\cc = A_\cc F_{\rm s} + F_{\rm b},\quad F_{\rm base} = F_{\rm s} + F_{\rm b},
\label{eqn:fcdef}
\end{equation}
\begin{equation}
t_{\rm r} = u_{\rm r} t_{\rm E} 
|\csc\phi|,\quad Q = F_{\rm s}^2 t_{\rm r},
\label{eqn:trdefs}
\end{equation}
and
\begin{equation}
\Delta t =  t_{\rm E}\rho_* |\csc\phi|.
\label{eqn:tstardef}
\end{equation}
Here $F_{\rm s}$ is the source flux, $F_{\rm b}$ is the background flux, $F_{\rm base}$
is the baseline flux, $A_\cc$ is the total magnification of the three
non-divergent images at the position of the caustic, and $u_{\rm r}$ characterizes
the square-root singularity of the caustic.  That is, in the neighborhood
of the singularity, the total magnification of the two divergent images is
given by $A_{\rm div}(u) = (\Delta u_\perp/u_{\rm r})^{-1/2}$, where 
$\Delta u_\perp$ is the
perpendicular distance from the position $u$ to the caustic in units of
the Einstein radius, $\theta_{\rm E}$.  
The Einstein crossing time is $t_{\rm E}=\theta_{\rm E}/\mu$, where $\mu$
is the relative source-lens proper motion, 
$\theta_{\rm E}^2 = (4 G M/c^2)(\dls/\dol\dos)$, $M$ is the total mass
of the binary, $\dol$ and $\dos$ are the distances to the lens and source,
and $\dls\equiv\dos-\dol$.  Finally, $t_{\rm r}$ is a parameter that 
characterizes the rise time of the caustic.

	In an ideal world, $t_{\cc}$, $Q$, $F_\cc$, and $F_{\rm base}$
would be measured exactly.  In this case, the search could be reduced to
four parameters $(d,q,\ell,t_{\rm E})$.  
(Recall that the ninth parameter, $\Delta t$, does not enter into the fit
to the non-caustic-crossing data.)
Here $d$ is the binary separation
in units of the Einstein ring, $q$ is the binary mass ratio, and $\ell$ is the
position along the caustic of the second caustic crossing.  For each pair
$(d,q)$, one steps along the caustic and determines $A_\cc$ and $u_{\rm r}$ from the
lens geometry.  Equation (\ref{eqn:fcdef}) yields
$F_{\rm s} = (F_\cc-F_{\rm base})/(A_\cc -1)$ 
and $F_{\rm b} = F_{\rm base} - F_{\rm s}$.  Next one
chooses a value of $t_{\rm E}$.  Equation (\ref{eqn:trdefs}) then fixes 
$\phi$: $|\sin\phi| = F_{\rm s}^2 u_{\rm r} t_{\rm E}/Q$.  This
completely determines the geometry and source trajectory (up to a two fold
ambiguity in $\phi$).  

	In practice, $Q$, $F_\cc$, and $F_{\rm base}$ all have
significant measurement uncertainties, and as Albrow et al.\ (1999c) 
discuss, this implies that one must allow a fifth continuous free parameter,
$\phi$, although this need be considered only over a restricted range.
If, as in the present case, light curve measurements come from several
telescopes in several bands, then
one must allow additional parameters for the source flux and background
flux for each.  However, these can be obtained from a simple linear fit
(see also below) and so do not add significant computation time.
The search through this 5-parameter space can be considerably simplified if
there is information about the time of first caustic crossing.  Then, for
each trial trajectory, one can first check if the last measured point before
the first caustic indeed lies outside the caustic and if the first point
after the first caustic indeed lies inside.  If either of these conditions
is not satisfied, the trajectory can be rejected without further investigation.

	For the second step of the method, we follow Albrow et al.\ (1999c)
with the following exceptions.  First, we include the
non-caustic-crossing data from all the light curves (except the SAAO $V$
band data which are too sparse to contain useful information for this
purpose).  

	Second, we restrict the first caustic crossing to lie between
the MPS point before the caustic at HJD$'=970.0503$ and the OGLE point
after the caustic at  HJD$'=970.9037$.  As noted in \S\ 2, the OGLE point
could in principle be on the {\it rising} side of the first caustic crossing,
i.e., {\it before} the center of the source crosses the caustic.  However,
a MACHO-CTIO $R$ band data point taken approximately
1.2 hours after the OGLE point rules out this possibility.
This point was not reported by Alcock et al.\ (1999a) due to an oversight 
but was reported by Rhie et al.\ (1999).  When the relative
normalizations of the different light curves are properly set, the MACHO-CTIO 
point lies $(20\pm 10)\%$ below the OGLE point.  If the OGLE point were
before the first caustic, we would expect the light curve to be rising
extremely rapidly, by of order a factor 2 in an hour, just as it is falling
very rapidly at the end of the second caustic crossing (Afonso et al.\ 1998).
Thus, the OGLE point certainly occurs on the falling side of the first caustic 
crossing.

	By restricting the first crossing to less than a day, we obtain 
a much more  powerful constraint
than the one used by Albrow et al.\ (1999c) who limited the first
caustic crossing only to HJD$'<973.8$.  However, this change
implies that smaller step sizes are required for $t_{\rm E}$ and
$\sin\phi$ so as to avoid missing the first caustic.  We choose 2\% increments
for each compared to 5\% used by Albrow et al.\ (1999c).  Since the two 
caustics are separated by 12 days, there are guaranteed to be at least
3 time steps for which the first caustic crossing lies in the designated 
range.  

	Finally, to avoid missing rotating wide binaries in the second step, 
we set the model
magnifications equal to unity ($A\equiv 1$) for all points prior to
HJD$'=810$ (i.e., about 160 days prior to the first caustic crossing).  
The MACHO data are fairly flat during this period (Fig.
\ref{fig:one}).  In fact, while many of the binaries that we consider
are at baseline during this early period, others, notably wide binaries,
are not.  They often show a ``bump'' (brightening then fading)
several hundred days before the caustic
crossing as the source approaches the companion star.  Since this bump is
not seen in the data, such binaries would seem to be ruled out.  However,
it is possible that the companion moved between the time that the source 
passed closest to the companion and the time when the source crossed the 
caustic (at which time the geometry of the event was primarily 
determined).
If it moved sufficiently far during this interval, then the source would
not have come close enough to the companion to cause a significant bump
(see \S\ 6.1).
Thus, we include the early data (since it helps set the baseline) but do
not allow it to rule out wide binaries until we have had a chance to 
examine the possibility that they might have avoided detection by rotating.

    From this step, we find two allowed regions in $(d,q)$ space.
One lies near $(d,q)\sim (0.5,0.5)$, and the other
near $(d,q)\sim (3.5,0.4)$.
Albrow et al.\ (1999a), Alcock et al.\ (1999a),
Udalski et al.\ (1998b), and Rhie et al.\ (1999) all considered solutions in
the general vicinity of the first region, but none considered solutions
near the second.  The two allowed regions lie in the lower right and upper
left quadrants of the broad range of possible solutions shown in Figure 6
from Albrow et al.\ (1999c).

	For the third step,  Albrow et al.\ (1999c) consider trial 
trajectories defined by seven parameters: the time 
$t_{\cc}$ and the duration $\Delta t$ of the caustic crossing, the Einstein
time scale $t_{\rm E}$, the projected separation of the binary
in units of the Einstein radius $d$,
the binary mass ratio $q=M_2/M_1$, 
the distance of closest approach (in units of the Einstein radius)
of the source to the midpoint of the 
binary, $u_0$, and
the angle $\alpha$ $(0\leq\alpha<2\pi)$
between the binary-separation vector ($M_2$ to $M_1$)
and the proper motion of the source relative to the origin of the binary.
(The center of the binary is taken to be
on the right hand side of the moving source.)\ \
 For each observation, the magnification is evaluated in one of
two ways.  If the source lies at least 3.5 source radii from the caustic, 
the magnification is simply the magnification of a point source.  If it is
closer, the finite size of the source is taken into account using the 
approximation
\begin{equation}
A^{\rm fs}({\bf u}_p) = A_3^0 ({\bf u}_p) 
+ A_2^0({\bf u}_q)\biggl({\Delta {u}_{q,\perp}\over\rho_*}
\biggr)^{ 1/2}
G_0\biggl(-{\Delta {u}_{p,\perp}\over \rho_*}\biggr),
\label{eqn:analytica}
\end{equation}
where ${\bf u}_p$ is the position in the Einstein ring of the center of the
source, 
and $\Delta {u}_{p,\perp}$ is the
perpendicular distances from ${\bf u}_p$ 
to the nearest caustic. 
If $\Delta {u}_{p,\perp}>\rho_*$, then ${\bf u}_q= {\bf u}_p$.
Otherwise, ${\bf u}_q$ is taken to lie along the perpendicular to the caustic
through ${\bf u}_p$ and halfway from the caustic to the limb of the star
that is inside the caustic.  The perpendicular distance from ${\bf u}_q$ to
the nearest caustic is $\Delta {u}_{q,\perp}$,
$A_3^0 ({\bf u}_p)$ is the magnification of the 3 non-divergent images
at the position ${\bf u}_p$, $A_2^0({\bf u}_q)$ is the magnification of the
2 divergent images at the position ${\bf u}_q$,
and $\rho_*$ is the source size in units of the Einstein ring.  
See Figure 3 from Albrow et al.\ (1999c).
The argument of $G_0$ is negative if the center 
of the source lies inside the caustic and positive if it lies outside.  
For each light curve $i=1, \ldots 14$, we then use standard linear techniques
to find the source flux $F_{s,i}$
and background flux $F_{b,i}$ that minimizes $\chi^2_i$,
\begin{equation}
\chi^2_i \equiv \sum_k{[F_{s,i}A^{\rm fs}(t_k) + F_{b,i} - F_k]^2\over \sigma_k^2},
\label{eqn:chi2def}
\end{equation}
where $F_k$ and $\sigma_k$ are the measured flux and error for the observation
at time $t_k$.  (We follow Albrow et al.\ 1999c in constraining $F_{s,i}$ to
be the same for the five PLANET light curves, $i=9,10,11,12,13$, and in
constraining $F_{b,10}=F_{b,13}$.)

\subsection{Limb-Darkening Parameterization}

	Equation (\ref{eqn:analytica})
is valid in the approximation that there is no limb darkening.
We model the surface brightness of the limb-darkened source by
\begin{equation}
S(\theta) = {F_{\rm s}\over \pi\theta_*^2}\biggl[1 - \Gamma\biggl[1-
{3\over 2}\biggl(1 - {\theta^2\over\theta_*^2}\biggr)^{1/2}\biggr]\biggr],
\label{eqn:surfprof}
\end{equation}
where $\theta$ is the angular position on the source star relative to its 
center, and $\Gamma$ is the limb-darkening parameter.  Note that with this 
formulation, there is no net flux in the $\Gamma$ term, so $F_{\rm s}$ remains
the total flux.  Convolving the $\Gamma$ term with the square-root singularity
of the caustic, we find the limb-darkened magnification is given by
\begin{equation}
A({\bf u}_p) = A^{\rm fs}({\bf u}_p) + \Gamma A^{\rm ld}({\bf u}_p)\qquad
 A^{\rm ld}({\bf u}_p) = A_2^0({\bf u}_q)
\biggl({\Delta {u}_{q,\perp}\over\rho_*}
\biggr)^{ 1/2}
H_{1/2}\biggl(-{\Delta {u}_{p,\perp}\over \rho_*}\biggr),
\label{eqn:analtyicld}
\end{equation}
where $H_n(\eta) \equiv G_n(\eta) - G_0(\eta)$, and
\begin{equation}
G_n(\eta)\equiv \pi^{-1/2}{(n+1)!\over (n+1/2)!}
\int_{\rm max(\eta,-1)}^{1} dx{(1-x^2)^{n+1/2} 
\over(x-\eta)^{1/2}}\,\Theta(1-\eta).\label{eqn:hdef}
\end{equation}
Explicitly
\begin{equation}
G_{1/2}(\eta) = {2\over 5}\sum_{\epsilon=\pm 1}
(3-2\epsilon\eta)(\epsilon -\eta)^{3/2}\Theta(\epsilon -\eta),
\label{eqn:Gonehalf}
\end{equation}
where $\Theta$ is a step function.
To allow for limb darkening, we then modify equation (\ref{eqn:chi2def}):
\begin{equation}
\chi^2_i \equiv \sum_k{[F_{s,i}A^{\rm fs}(t_k) + F_{ld,i}A^{\rm ld}(t_k) +
F_{b,i} - F_k]^2\over \sigma_k^2}.
\label{eqn:chi2defmod}
\end{equation}
The limb-darkening parameter for light curve $i$ is then 
$\Gamma_i = F_{ld,i}/F_{s,i}$.

	It is conventional to parameterize limb darkening by 
\begin{equation}
S(\theta) = S(0)\biggl[1 - c\biggl[1-
\biggl(1 - {\theta^2\over\theta_*^2}\biggr)^{1/2}\biggr]\biggr].
\label{eqn:surfprofcov}
\end{equation}
In this case, the flux associated with the limb-darkening term is not zero.
Rather, it is a fraction $(3/c -1)^{-1}$ of the total flux.  In a 
multi-parameter problem, the limb-darkening parameter then develops 
correlations
with other parameters with which it has no physical connection.  In our
formulation, there is no net flux in the limb-darkening term, so the effect
of limb darkening rapidly and explicitly vanishes far from the caustic 
crossing,
\begin{equation}
H_{1/2}(\eta) \rightarrow -{3\over 160}(-\eta)^{-5/2},\qquad(\eta\ll -1).
\label{eqn:Honehalflim}
\end{equation}
Thus there are no spurious correlations.  To make contact with the usual
formulation, we note that
\begin{equation}
c = {3\,\Gamma\over \Gamma + 2}.
\label{eqn:cequivgamma}
\end{equation}

	Limb darkening affects the magnification only if the source is
transitting or is very close to the caustic.  Thus, in principle it could
affect the SAAO $V$ and $I$ curves (which both covered most of the caustic
crossing), the Yale-CTIO curve (which has one point just before the end of
the caustic crossing), the EROS $B$ and $R$ curves  (which have 16
points each during the last 110 minutes of the caustic crossing),
the MACHO CTIO $R$ curve (which has 2 points just before the end
of the crossing), and the MACHO $B$ and $R$ curves (which have points up
to $1.7\Delta t$ {\it before} the caustic crossing).

	While the Yale-CTIO curve does not
have enough coverage of the caustic crossing to make an independent estimate
of the limb darkening, it is tied to the SAAO photometry
as discussed following equation (\ref{eqn:chi2def}) and more thoroughly
in \S\ 2 of Albrow et al.\ (1999c).  Thus, this one 
Yale-CTIO point can enter the fit for the SAAO $I$ limb-darkening parameter.
On the other hand, from equation (\ref{eqn:Honehalflim}),
we see that limb darkening affects the MACHO $B$ and $R$ fluxes by less
than a fractional amount 
$\Gamma H_{1/2}(\eta)/G_0(\eta)\sim (3/160)\eta^{-2}\Gamma\la 0.3\%$,
where we have assumed $\Gamma\la 0.5$ and where we have made use of the
limiting form $G_0(\eta)\sim (-\eta)^{-1/2}$ for $\eta\ll -1$.  This
compares to typical errors in MACHO photometry for these exposures 
of 2\% to 3\%.  Hence we do not attempt to fit limb-darkening
parameters to these two light curves.  Thus, we fit for five independent
limb-darkening parameters: SAAO $V$, EROS $B$, MACHO CTIO $R$,
EROS $R$,  and SAAO $I$, with corresponding central wavelengths of
0.55, 0.62, 0.64, 0.76, and 0.80 $\mu$m.

\section{Static-Binary Solutions}

	We first search for solutions with static binaries.  To do so, 
we will again set $A\equiv 1$ for all points with
HJD$'<810$.  In \S\ 6, we will then investigate whether the solutions found
in this way (or solutions near them) are in fact permitted when binary
rotation is taken into account.  We conduct the search on a grid with
$(\Delta d,\Delta q)=(0.02,0.02)$ for the close-binary solution and
$(\Delta d,\Delta q)=(0.05,0.04)$ for the wide-binary solution.

	We find two sets of static solutions.  One is centered 
at $(d,q)=(0.54,0.50)$.  At the $3\,\sigma$ level $(\Delta\chi^2<9)$,
it extends from about  $(d,q)=(0.46,0.42)$ to about $(d,q)=(0.60,0.58)$,
and is about half as wide in the orthogonal direction.  The other solution
is centered at  $(d,q)=(3.25,0.24)$ and at the $3\sigma$ level
extends over the range $d=3.25^{+0.40}_{-0.20}$ and 
$q=0.24^{+0.20}_{-0.16}$.  
Dominik (1999b) has argued that there is a generic degeneracy
in fitting light curves between a  pair of close-binary and wide-binary
solutions.  The second (wide-binary) solution
is formally favored at the $2\,\sigma$ level ($\Delta\chi^2=4$), but we 
do not consider this to be a compelling reason to adopt it as the preferred 
solution.

All the solutions near the close-binary minimum have similar parameters,
as do all the solutions near the wide-binary minimum.  For 
simplicity we quote the full set of parameters only at the minimum.  
These are shown in Tables 1 and 2.  The division between the two tables
is such that the parameters shown in Table 2 are derived from the
linear fit described by equation (\ref{eqn:chi2defmod}) and so have
associated error bars.  The remaining parameters are shown in Table 1.
Note that only the first 7 parameters in Table 1 are independent.  The
five remaining parameters are derived from the fit.  In particular,
$t_* = \Delta t \sin\phi$, and 
$t_0= t_\cc - t_{\rm E}(u_{\cc,x}\cos\alpha +u_{\cc,y}\sin\alpha)$.
We caution that 
the numbers of decimal places given for 
the parameters in Table 1 convey a much higher precision than the
statistical errors (which are in fact not even precisely known).  The
purpose of presenting many decimal places is to allow the reader to
reproduce the solution.  Because of strong
correlations among the parameters, their values in a particular model must
be known to high precision in order to avoid misdirecting the model into
inappropriate regions of parameter space.
The error bars in Table 2 reflect only the correlations within the linear
fit described by equation (\ref{eqn:chi2defmod}) and not the correlations
with the parameters in Table 1.  Therefore these are actually lower limits
on the errors.  Note that we show the ratio $F_{\rm b}/F_{\rm s}$ only for the
light curves for which it is reasonably well determined ($<10\%$).

The parameter that varies the most over the allowed set of the solutions
is $t_{\rm E}$, which ranges from about 75 to about 125 days within
the $3\,\sigma$ range of the close-binary solutions and from about
145 to 200 days within the $3\,\sigma$ range of the wide-binary solutions.  
 From the
standpoint of the proper-motion measurement, three parameter combinations 
are important, $(V-I)_{\rm s}$, $I_{\rm s}$, and $t_*=\Delta t \sin\phi$.  The first
essentially does not vary at all, $(V-I)_{\rm s}=0.30$ for all allowed solutions.
When $(V-I)_{\rm s}$ is fixed, the proper motion scales as
$\mu\propto 10^{-0.2 I_{\rm s}}t_*^{-1}$.  The full ($3\,\sigma$)
range of variation of
this parameter combination and thus of $\mu$ is only about 25\% over each
the two classes of solutions.

The limb-darkening coefficients given in Table 2 are shown 
in Figure \ref{fig:limb}.  The close-binary is shown by open circles and
the wide-binary is shown by filled circles.  The
horizontal error bars show the full-width at half maximum (FWHM)
of the filters,
while the vertical error bars denote the statistical errors.  We 
emphasize again, however,
that these include only the errors from the linear fit generated by equation
(\ref{eqn:chi2defmod}) and not those that arise from correlations with
the other parameters.  If we suppress limb darkening and force a fit to
a uniform disk, then in both cases,
$\chi^2$ increases by about $\Delta\chi^2=38$ for 5 degrees
of freedom.  This is far less than the $\Delta\chi^2=106$ that would be
predicted based on a naive interpretation of the errror bars shown
in Figure \ref{fig:limb}, and
this
difference arises exactly from the fact that these error bars 
do not account for
the correlations with other parameters.  Nevertheless, the full fit
reveals that limb darkening has been detected with high significance
(formally 99.999\%).
%

	Unfortunately, to the best of our knowledge, no theorists have ever
calculated limb-darkening profiles of metal-poor A stars.  Since limb
darkening has clearly been detected in one such star, perhaps some
theorist will now undertake such a calculation.  For the close-binary
solution, the limb-darkening
coefficients fall from $0.45\pm 0.11$ for $V$ to
$0.17\pm0.04$ for $I$.  The wide-binary solution is similar.
The one exception is MACHO-CTIO $R$, but its
error bars are too large to make any definite
statement because its limb-darkening parameter was derived from only two
measurements.

	Tables 1 and 2 also show a third solution, one for a 
{\it rotating} wide binary.  This solution is derived in \S\ 6.1, below.
As is clear from Figure \ref{fig:one}, the static wide-binary solution is
not viable: it is only an intermediate step on the way to finding a
viable rotating wide-binary solution.  Hence, in Tables 1 and 2, the
close-binary and rotating wide-binary solutions are labelled ``viable''
while the static wide-binary solution is labelled ``not viable''.
However, until we introduce rotation in \S\ 6, all references to the
wide-binary solution will be to the {\it static} version.

	Figures \ref{fig:two} and \ref{fig:three} show the model light curves
together with all the available data for the close-binary and wide-binary
solutions respectively.  Because the data are in different passbands, we
cannot compare the predicted flux with the observed flux as we could if the
data were in a single passband.  We therefore deblend our data, i.e.\ we
plot $2.5\,\log[(F-F_{\rm b})/F_{\rm s}]$ (points) and compare this to
$2.5\,\log$(magnification) (solid curve), 
where $F_{\rm s}$ and $F_{\rm b}$ are the fit values of the source and
background flux.  The points are binned primarily in one-day intervals.
However, the points before HJD$'=950$ are binned in 10-day intervals and
the points near the caustic crossings are binned in 0.1-day intervals.
Data from different observatories are combined together.
Figures \ref{fig:four} and \ref{fig:five} show close-ups of the two
model fits in the neighborhood of the second caustic crossing binned in
0.01-day intervals.  

	The two fits appear to be equally good to the eye. 
This is illustrated in Figure \ref{fig:six}
which shows the fractional difference in the predicted fluxes between the
two models for each of the 14 light curves analyzed in this paper.  The
fundamental physical reason for this degeneracy is shown in Figure 
\ref{fig:seven} where the caustic structures for the two solutions are
superposed.  These caustic structures are very similar.

\section{Proper Motion}

	The proper motion is given by $\mu=\theta_*/t_*$.  To obtain the
proper motion one must therefore estimate the angular source size $\theta_*$,
which can be calculated if one knows the dereddened color and magnitude of the
source.  Among all the light curves, there are photometric
calibrations for only five: PLANET $V$ (SAAO only)
and PLANET $I$ (SAAO and Yale-CTIO) (Albrow et al.\ 1999a),
MACHO $B$ and  $R$ (Alcock et al.\ 1999b), and OGLE $I$.  As we describe
below, the calibration of PLANET $I$ is tied to the OGLE calibration.
We find that the $F_{\rm s}$ values for these two light curves are consistent at
the $1\,\sigma$ level.  However, the errors for the OGLE $I$ $F_{\rm s}$ are an order
of magnitude larger than for PLANET $I$ (because there are many fewer data
points), so the OGLE $I$ $F_{\rm s}$ does not yield significant additional 
information about the flux of the source.
For the close-binary solution, we have
$V_{\rm s} = 22.36\pm 0.01$,
$I_{\rm s} = 22.06\pm 0.00$,
$(V-I)_{\rm s}=0.30\pm 0.01$,
from PLANET and 
$V_{\rm s} = 22.67\pm 0.01$,
$R_{\rm s} = 22.58\pm 0.01$,
$(V-R)_{\rm s} = 0.09\pm 0.01$ from MACHO.
For the wide-binary solution, we have
$V_{\rm s} = 22.17\pm 0.01$,
$I_{\rm s} = 21.87\pm 0.00$,
$(V-I)_{\rm s}=0.30\pm 0.01$,
from PLANET and 
$V_{\rm s} = 22.48\pm 0.01$,
$R_{\rm s} = 22.39\pm 0.01$,
$(V-R)_{\rm s} = 0.09\pm 0.01$ from MACHO.

	In addition to these errors reported by the fit, Albrow et al.\ (1999a)
estimate that their calibration error is 0.02 mag for the PLANET color and
Alcock et al.\ (1999b) estimate that their calibration error is 0.04 mag for
the MACHO color and 0.10 for the magnitudes.  Two points are clear from this
summary.  First, the {\it ratios} of fluxes are essentially identical for
the two models.  Second, the MACHO and PLANET colors are mildly inconsistent
and the MACHO and PLANET $V$ magnitudes are inconsistent at the $3\,\sigma$
level.  We believe
that the PLANET calibration is substantially more reliable than the MACHO
calibration since PLANET calibrated their data using secondary standards in the
field that were in turn measured in the standard way by OGLE (Udalski et al.\
1998a), i.e. from primary standards on photometric nights.  On the other hand,
although MACHO applies essentially the same procedure for their calibration
of their LMC fields, for the SMC they simply adopt the mean zero points
derived for the LMC at similar airmass (Alcock et al.\ 1999b).  We therefore
adopt the PLANET calibration.

	Following Albrow et al.\ (1999a), we adopt a total extinction of
$A_V=0.22\pm 0.1$.  The final
results do not depend strongly on the extinction (see below).  
The flux is given by $F= \theta_*^2 S$, where $S$ is the
mean surface brightness of the source.  We will assume that this surface
brightness is a function only of the $(V-I)_0$ color and not any other
properties of the star.   (We know, for example, that the star is a dwarf
rather than a giant.)\ \ We can then write
\begin{equation}
\theta_* = 
79\,{\rm nas}\,
10^{-0.2(I_0-22)}\biggl({S\over S_{(V-I)_0=0.21}}\biggr)^{-1/2}
\label{eqn:surfbrite}
\end{equation}
where we have evaluated the normalization using Green, Demarque, \& King 
(1987), specifically their $Y=0.2$, $Z=0.001$, Age = 1 Gyr table.  We therefore
obtain estimates of 82 nas and 89 nas for the angular size of the source in 
the close-binary and wide-binary solutions respectively.
As described by Albrow et al.\ (1999a), this estimate has a 3\% error for
uncertainty in the extinction $A_V$ (Albrow et al.\ 1999a) 
and a 5\% error for uncertainty in the theoretical model
(M. Pinsonneault 1998, private communication), 
for a total uncertainty of 6\%.

	Hence in the two models the proper motions are
\begin{equation}
\mu = 1.30\pm 0.08\,\kms\,\kpc^{-1} \qquad ({\rm close}\ {\rm binary})
\label{eqn:mod1}
\end{equation}
and
\begin{equation}
\mu = 1.76\pm 0.11\,\kms\,\kpc^{-1} \qquad ({\rm wide}\ {\rm binary})
\label{eqn:mod2}
\end{equation}
The errors in these equations reflect only the uncertainties in the 
extinction and the stellar models, and they do not include uncertainties in
the parameter fits.  Recall from \S\ 4, however, that even at the $3\,\sigma$
level, the range of allowed values of the parameter combination
$10^{-0.2 I}t_*^{-1}\propto \mu$ is very restricted.

	The values in equation (\ref{eqn:mod1}) and (\ref{eqn:mod2})
clearly put the lens in the SMC rather than the Galactic
halo.  For comparison note that Albrow et al.\ (1999c) found some
solutions that were moving much faster and hence would not be 
explainable
as SMC events.  These additional solutions are ruled out by combining all
the available data.

\subsection{Binary Physical Characteristics}

	Since the binary is known to be in the SMC, we can use the
proper-motion measurements to obtain estimates of the binary physical
projected separation, 
\begin{equation}
r_{\rm p}=d\mu\, t_{\rm E}\dos \simeq d\mu t_{\rm E}\times 60\,\kpc.
\label{eqn:projsep}
\end{equation}
This yields $r_{\rm p}=2.40$ AU and $r_{\rm p}=32.4$ 
AU for the close-binary and wide-binary
solutions respectively.  Note that for circular face-on orbits, $r_{\rm p}=a$,
the semi-major axis, while all orbits satisfy $a>r_{\rm p}/2$.

	These projected separations can be used to place limits on
the motion of the binary.  For example, for the close-binary
solution, the
blended background flux in the MACHO $B$ band light curve (which has the
best-determined blended flux and is also the only well-determined measurement
in the blue) is approximately 50\% larger than
the source flux, so the larger of the two lens stars cannot be more
than about $2.5\,M_\odot$.  For the wide-binary solution, the blended and
source fluxes are about equal, so the larger star cannot be more than
about $2\,M_\odot$.
Thus the total mass of the binary in both cases is limited to
$M\la 3\,M_\odot$.
If we momentarily assume a face-on circular
orbit, then from Kepler's Third Law, the period is constrained to
$P_{\rm }>2\,$yr and $P_{\rm }>110\,$yr for the two solutions. 
 For a face-on eccentric orbit at apastron, the
periods could actually be shorter by $8^{1/2}$, but what actually concerns us
is not the length of the period but the relative motion of the binary lenses
over times that are very short compared to the period.  For a circular
orbit, this instantaneous angular speed is 
$\omega_{\rm circ} = 2\pi/ P_{\rm circ}$, but the
maximum instantaneous angular speed occurs for a face-on eccentric orbit 
where the caustic crossing occurs near periastron:
$\omega_{\rm max} = 2^{1/2}\omega_{\rm circ}$.  We must therefore consider
binary motions up to this level.

\section{Rotating Binaries}

	Although binaries are not static, only a few attempts have
been made to fit microlensing light curves to dynamic binaries (Dominik 1998).
In principle, it is possible to measure six orbital parameters
of a binary from sufficiently precise observations.  These are the
same six that can be measured from proper-motion measurements of visual
binaries except that the angular semi-major axis is measured relative
to $\theta_{\rm E}$ (rather than absolutely) and the line of nodes is
measured relative to the direction of the source (rather than celestial
coordinates).  In practice, it is extremely difficult to measure anything
other than the (two-dimensional) projected relative velocity of the
components in units of $\theta_{\rm E}$.  In fact, no binary-motion
information of any type has ever been extracted from a microlensing event.
We therefore restrict consideration to the simplest form of such motion,
uniform circular motion in the plane of the sky.  
This leaves the geometry of the lens fixed and
permits only rotation of this geometry.  If we allowed more general
two-dimensional motion, the geometry of the lens would change as the
projected positions of the two components moved closer together or further 
apart.  We will explicitly
ignore this type of change in the binary configuration.

\subsection{Wide-Binary Solution}

	As we discussed in \S\ 3, we forced the magnification at early
times to $A=1$ when fitting the static binary solutions because the
MACHO light curve is observed to be flat at these times.  Had we not
done so, the wide binary solution would have been ruled out at the $18\,
\sigma$ level
($\Delta\chi^2=342$).  
In Figure \ref{fig:one} we show the early light curve for the 
best-fit static-binary solution together with the MACHO data.  The model
is clearly ruled out by the data.  In fact, we find that even if we allow
this binary to rotate with a period of $P=75\,$yr, i.e., the minimum 
permitted by the argument of \S\ 5.1, the model is still
ruled out at the $8\,\sigma$ level ($\Delta\chi^2=58$).  However, there
are satisfactory
rotating binaries in the neighborhood of the best-fit static solution.
In Tables 1 and 2 we give the parameters for one of these
rotating solutions with $(d,q)=(3.65,0.36)$ and $P=75\,$yr, and in Figure
\ref{fig:eight} we show a diagram of the caustic structure for this 
rotating solution together 
with the corresponding static solution for the same $(d,q)$.  
Since the two lenses are separated by much more
than an Einstein ring, the magnification structure is for the most part
a superposition of the magnification of two isolated lenses.  Hence, it
is clear from the diagram why the static
model is excluded: the source passes within
$\sim 0.4$ binary-mass Einstein radii of the larger lens, which is about
0.55 Einstein radii scaled to the mass of this lens.  Thus, the magnification
is about 2.  Even though the event is heavily blended ($F_{\rm b}/F_{\rm s}\sim 2$) and
the errors in MACHO photometry are relatively large at these early times,
this magnification would easily be seen in the data.  However, if the
binary were rotating clockwise in the plane of the sky, 
then two years before the caustic crossing
at the time of closest approach, the source would have been about twice as
far from the heavier lens, thus reducing $(A-1)$ by a factor 3.5.
We show the light curve resulting from this rotating-binary solution
in Figure \ref{fig:one}. It is barely distinguishable from
the baseline.  Numerically we find that the 75-year period
binary increases $\chi^2$ by less than 1 unit relative to the artificial case
used in the initial simulations of $A\equiv 1$ for HJD$'<810$.  We find that
the $\chi^2$ of this rotating wide binary is only 5 higher than the $\chi^2$
of the best-fit close binary.
We conclude that the data are consistent with a wide-binary solution.
The upper panel of Figure 
\ref{fig:eight} is a close up of the trajectories with and without rotation.
The two trajectories are essentially identical in the region
around the caustic crossing.  

	Because the allowed rotating binaries tend to be on one side of
best-fit static wide-binary solution (higher $d$ and higher $q$), they
tend to have systematically lower proper motions than that of the static
solution given in Tables 1 and 2.  For the rotating wide binary shown in
Figure \ref{fig:eight} (and indeed for its static analog), we find
\begin{equation}
\mu = 1.48\pm 0.09\,\kms\,\kpc^{-1} \qquad 
({\rm rotating}\ {\rm wide}\ {\rm binary}).
\label{eqn:mod3}
\end{equation}

\subsection{Close-Binary Solution}

	For the close-binary solution the rotation periods can be much shorter,
so that in contrast to the situation illustrated in Figure \ref{fig:eight}
for the wide binary, the rotating and non-rotating trajectories are not the
same within the caustic region.  Hence, rotating solutions require 
substantially different geometries.  For example, for $P=10$ yr 
(counter-clockwise), we find a best fit at 
$(d,q,t_{\rm E}) = (0.56,0.44,95.5$ days), and
for $P=10$ yr (clockwise), we find
$(d,q,t_{\rm E}) = (0.54,0.58,95.9$ days). 
These solutions are
about $1\,\sigma$ worse fit than the non-rotating solutions.
Their proper motions are respectively 4\% lower and 7\% higher,
and the limb-darkening parameters are very similar to the non-rotating case.
Thus, while both rotating and non-rotating solutions are compatible with the
data and while allowing rotation increases the uncertainties in the
binary parameters, these various solutions 
have very similar implications for the nature of the source and lens.

\section{Comparison With Previous Solutions}

	Six papers have made estimates of some or all of the parameters
of MACHO 98-SMC-1 based on subsets of the data presented here
(Afonso et al.\ 1998; Albrow et al.\ 1999a,c; Alcock et al.\ 1999a; 
Udalski et al.\ 1998b; Rhie et al.\ 1999).  We now analyze the relationship
of the results presented in this work to these earlier efforts.  We will
focus attention on whether the various previous solutions and partial
solutions are consistent with one another and with the present results,
and we will attempt to resolve any inconsistencies.

	Rhie et al.\ (1999) analyzed almost all of the non-PLANET data
presented here, and Albrow et al.\ (1999c) analyzed almost all of the PLANET
data.  Hence, our present analysis is essentially based on the union of
these two disjoint data sets.  

	Albrow et al.\ (1999a) published two solutions, now known as
PLANET Model I and PLANET Model II.  
However, Albrow et al.\ (1999c) subsequently showed that PLANET Model I
actually sits in a {\it single} extremely broad, 
virtually flat, $\chi^2$ minimum which connects {\it all} of the 
close-binary solutions that they found.  PLANET Model I is essentially
the same as model \#26 from Albrow et al.\ (1999c).  
Because of their excellent coverage of the
second caustic crossing, Albrow et al.\ (1999c) were able to measure
$t_\cc$ very precisely and $\Delta t$ fairly precisely.  
Those measurements are confirmed
by the solutions presented here.  On the other hand, they showed that the
broad degeneracy in their overall solution could be traced to their lack
of coverage of the early light curve (see their Figs.\ 7, 8, and 9).
One would expect as more data are added to the data available to
Albrow et al.\ (1999c), that the two broad minima shown in their Figure
6 would contract and possibly break up into several discrete local minima.
The close-binary solution presented here is very similar to
an interpolation between the neighboring grid points of their
models \#27 and \#31.

	Afonso et al.\ (1998) measured the parameter combination
$t_\cc+\Delta t=982.8039\pm 0.0010$ (after correction of a transcription 
error in the original paper) based on EROS coverage of the
end of the light curve.  This differs by only 2 minutes from the values shown
in Table 1.  Alcock et al.\ (1999a) modeled the event by combining 
their own MACHO/GMAN data with
the EROS data.  The MACHO/GMAN model was refined by Rhie et al.\ (1999) 
after the time
of the first caustic was pinned down by their own MPS data together with
the OGLE data (Udalski et al.\ 1998b).  We now investigate the consistency
of the MPS model with the models based solely on the PLANET data on the one
hand, and with the close-binary (CB) model presented here on the other.

	The first question to ask is: are the MPS and CB models in discrete
local minima, or are they two different points in the same minimum?
They are located at $(d,q,t_{\rm E})=(0.646,0.518,70.5)$ and
$(0.54,0.50,99.0)$ respectively.  To answer this question, 
we find solutions based
on all the data, but subject to the constraint of fixed $(d,q)$. We
evaluate these solutions on a grid of $(\Delta d,\Delta q)=(0.02,0.02)$
in the neighborhood of the CB model at $(d,q)=(0.54,0.50)$.
We find that $\chi^2$ varies smoothly over this grid of solutions.
The solution at $(d,q)=(0.64,0.52)$ is extremely similar to the MPS solution,
and $\chi^2$ rises monotonically between the CB and MPS-like solutions
$(\Delta\chi^2=32)$.  Hence these two solutions are in the same minimum
and are not discrete minima.

	Since the MPS solution has higher $\chi^2$ based on all the data
and is in the same minimum as the CB solution, did MPS therefore find a false 
minimum?  To address this we evaluate $\chi^2$ for the two solutions based
on the data available to MPS (i.e.\ excluding the PLANET data and the EROS data
from other than the night of the caustic crossing and using the SoDoPhot
reductions of the MACHO data rather than image subtraction.)  We then find that
the MPS solution is favored over CB by $\Delta\chi^2=10$.  That is, the
two solutions differ because they are based on different data sets rather
than because of different modeling procedures.

	Finally, we investigate a conflict between the MPS and PLANET
solutions which was previously identified by Rhie et al.\ (1999).  They
noted that their value of $t_\cc=982.683$ or $t_\cc=982.694$ 
(depending on their model of limb darkening) is later than the PLANET
value $t_{\cc} = 982.62439\pm 0.00087$ (Albrow et al.\ 1999c and confirmed 
here).  They did not quote error bars on their own value which is based
on modeling the interpolation between MACHO data cutting off 0.3 days before
the crossing and EROS data beginning 0.1 days after it.  However, by
evaluating $\chi^2$ for a series of models with the parameters
$d,q,\alpha,u_0,
t_{\rm E}$, and $t_\cc+\Delta t$ fixed, but $t_\cc$ varying,
we find that the error in the MPS value for $t_\cc$
is approximately 0.009 days.  Thus, the difference
between the MPS and PLANET values is a $6\,\sigma$ discrepancy if due
to an MPS problem and a $68\,\sigma$ discrepancy if due to a PLANET problem.
Clearly this difference is not the product of a statistical fluctuation.

	To determine the origin of this conflict, we fit all the data
but use the original SoDoPhot reductions (used by MPS) in place
of the image-subtraction reductions (used here) for the MACHO data.  We find
that the MACHO data points lie systematically above the model at
the beginning of night before the caustic crossing (HJD$'\sim 982.1$) and
systematically below the model at the end of the night 
(HJD$'\sim 982.3$).  There is no such systematic trend in the MACHO
data points when reduced by image subtraction.  We infer that this trend
may have been responsible for the late $t_\cc$ in the MPS model.  We
test this hypothesis by redoing the fits based on an MPS-like data set
but substituting image-subtraction reductions for SoDoPhot.  We then find
that the $(d,q)=(0.54,0.50)$ solution is favored over the $(d,q)=(0.64,0.52)$
solution by $\Delta\chi^2=5$.  Moreover the $t_\cc$ in the best-fit model
now differs from the PLANET value by only 0.02 days which is only a
$2\,\sigma$ discrepancy.

	In brief, the EROS measurement of $t_\cc+\Delta t$ and the PLANET
measurement of $t_\cc$ have been confirmed with high precision.  The
original MACHO model when refined by MPS based on the MPS+OGLE
determination of the first caustic crossing holds up very well.  It
lies close to the CB solution based on all the data.  In hindsight,
image subtraction would have yielded even more precise refinements of this
model.  Finally,
one sub-region of the broad class of wide-binary solutions found by
PLANET (Albrow et al.\ 1999c) survives the inclusion of the non-PLANET data.
We pat our collective selves on the back for a job well done.

\section{Conclusions}

	We have combined the data on MACHO 98-SMC-1 from five collaborations
to produce one of the best sampled microlensing light curves ever published.
We confirm earlier claims that the relative source-lens proper motion is low,
so the lens must be in the SMC.  However, there is a twist:
despite the fact that our combined data set is enormously superior to any
of the individual data sets, there are two very distinct solutions that
are compatible with all the data, a close-binary and a wide-binary solution.
Fortunately, both have very similar proper motions so there is no significant
ambiguity in this parameter.
We find a relative proper motion of $\mu\sim 1.30\,\kms\,\kpc^{-1}$ or
$\mu\sim 1.48\,\kms\,\kpc^{-1}$.

	We have measured the limb-darkening parameter in five different bands
with centers at 0.80, 0.76, 64, 0.62, and 0.55 $\mu$m.  If our results
are expressed in terms of the standard limb-darkening parameter $c$, the
respective values for the close-binary solution are
$0.23 \pm 0.05$,
$0.24 \pm 0.05$,
$0.06 \pm 0.34$,
$0.42 \pm 0.05$, and
$0.55 \pm 0.11$.
All other solutions have limb-darkening parameters that are close to these.

\begin{acknowledgements}
The EROS collaboration wishes to thank J.F.\ Lecointe for assistance
with the online computing, the ESO staff at La Silla Observatory, and
the Observatoire de Haute Provence.
Work by the MACHO/GMAN collaboration was supported by
DOE contract W7405-ENG-48 and DOE grant DEF03-80-ER 40546, 
NSF grant AST-8809616, a 
grant from the Bilateral Science and
Technology Program of the Australian Department of Industry,
Technology and Regional Development, a PPARC Advanced Fellowship,
a Packard Fellowship, and Fondecyt 1990440.
Work by the MPS collaboration was supported by
a grant from the NASA Origins 
program (NAG5-4573), the National Science Foundation (AST96-19575), 
by a Research Innovation Award from the Research Corporation,
by a grant from the Bilateral Science and
Technology Program of the Australian Department of Industry,
Technology and Regional Development, and  
by a grant from the Office of Science and Technology
Centers of NSF under cooperative agreement AST-8809616
Work by the OGLE collaboration was supported by
Polish KBN grant 2P03D00814 and NSF grant AST-9530478.
Work by the PLANET collaboration was supported by
grants AST 97-27520 and AST 95-30619 from the NSF, 
by grant NAG5-7589 from NASA, by grant ASTRON 781.76.018 from the
Dutch Foundation for Scientific Research (NWO), and
by a Marie Curie Fellowship (grant ERBFMBICT972457)
from the European Union.  PLANET wishes to thank the observatories that 
support its science, Canopus, CTIO, Perth, and SAAO, for the large awards
of telescope time that make intense microlensing monitoring possible.

\end{acknowledgements}

\newpage

\begin{figure}
\caption[junk]{\label{fig:one}
MACHO $B$ and $R$ data  for MACHO 98-SMC-1 binned in 20-day intervals for
the period before HJD$'=810$.  The bold lines are the values for 
the baseline flux from the close-binary solution, $(d,q)=(0.54,0.50)$.
 The event shows no significant deviation
from baseline during this early period.  The solid lines are from the best 
fit for
the {\it non-rotating} wide binary $(d,q)=(3.25,0.24)$ which is clearly ruled
out by the data.  However, a wide binary with a 75 year period and a 
nearby $(d,q)=(3.65,0.36)$ is permitted.  Its early light curve is 
shown as a dashed curve that
is barely distinguishable from a flat-baseline curve.
}
\end{figure}

\begin{figure}
\caption[junk]{\label{fig:limb}
Limb-darkening parameters $\Gamma$ (as defined in eq.\ \ref{eqn:surfprof})
derived from the close-binary solution $(d,q)=(0.54,0.50)$ ({\it open circles})
and the wide-binary solution $(d,q)=(3.25,0.24)$ ({\it filled circles})
to MACHO 98-SMC-1
for 5 passbands.  From left to right: PLANET-SAAO $V$, EROS $B$, 
MACHO-CTIO $R$, EROS $R$, and PLANET-SAAO $I$.  The horizontal error bars
represent the full-width at half-maximum of the filters, and the vertical
error bars are statistical.  The limb-darkening parameters for the 
two solution are very similar because the measurement of limb
darkening depends primarily on the caustic crossing and not on the global
characteristics of the light curve.   To avoid clutter, the error bars 
for the wide-binary solution are not shown, but they are almost identical
to the error bars for the close-binary solution.
}
\end{figure}

\begin{figure}
\caption[junk]{\label{fig:two}
Predicted versus ``observed'' {\it deblended}
magnification for the close-binary model $(d,q)=(0.54,0.50)$.
The deblended 
magnification is $A= (F-F_{\rm b})/F_{\rm s}$ where $F$ is the observed
flux, and $F_{\rm s}$ and $F_{\rm b}$ are the fit source and background fluxes in the
model.  Data are binned, mostly in 1-day bins.  However, for HJD$'<950$ there
are 10-day bins, and in the immediate neighborhood of the caustics there
are 0.1-day bins.  Data from all 14 light curves from the 5 collaborations
are averaged together whenever they lie sufficiently close to fit in the
same bin.
}
\end{figure}

\begin{figure}
\caption[junk]{\label{fig:three}
Predicted versus ``observed'' {\it deblended}
magnification for the wide-binary model $(d,q)=(3.25,0.24)$.
Similar to Fig.\ \ref{fig:two}.
}
\end{figure}

\begin{figure}
\caption[junk]{\label{fig:four}
Predicted versus ``observed'' {\it deblended}
magnification for the close-binary model $(d,q)=(0.54,0.50)$
showing the vicinity of the second caustic crossing.  Same as Fig.\ 
\ref{fig:two} except that bins are 0.01 days.
}
\end{figure}

\begin{figure}
\caption[junk]{\label{fig:five}
Predicted versus ``observed'' {\it deblended}
magnification for the wide-binary model $(d,q)=(3.25,0.24)$
showing the vicinity of the second caustic crossing.  Same as Fig.\ 
\ref{fig:three} except that bins are 0.01 days.
}
\end{figure}

\begin{figure}
\caption[junk]{\label{fig:six}
Fractional differences between fluxes predicted by 
the close-binary solution, $(d,q)=(0.54,0.50)$, and 
the wide-binary solution, $(d,q)=(3.25,0.24)$, for the 14 different light
curves ({\it solid lines}).
}
\end{figure}

\begin{figure}
\caption[junk]{\label{fig:seven}
Caustic structures for the close-binary ({\it bold curve}) and wide-binary
({\it solid curve}) solutions.  Each has been rescaled according to
the Einstein crossing time of the solution.  The caustics have been rotated
so that the source trajectories ({\it straight solid line}) overlap.
Time is shown in days from the second caustic crossing, so source motion
is to the right.
}
\end{figure}

\begin{figure}
\caption[junk]{\label{fig:eight}
Wide-binary trajectory with $(d,q)=(3.65,0.36)$
for static case and for binary with $P=75$ yr period.
The upper panel shows a close-up of the caustic together with the two
source trajectories which are barely distinguishable.  The light curve in
this region is therefore independent of rotation and the structure of the
caustic fixes the local trajectory.  The lower panel shows the full caustic
structure.  The two caustics are separated by about 3.65 Einstein radii, or
about 2 years.  This interval
is sufficient to allow the source-companion closest approach to grow by
by a factor $\sim 2$ relative to the static case.  This in turn reduces
$(A-1)$ a factor of $\sim 3.5$.  The static solution is ruled
out by the data but the rotating solution is permitted
(Fig.\ \ref{fig:one}).  The source is moving to the left.
}
\end{figure}


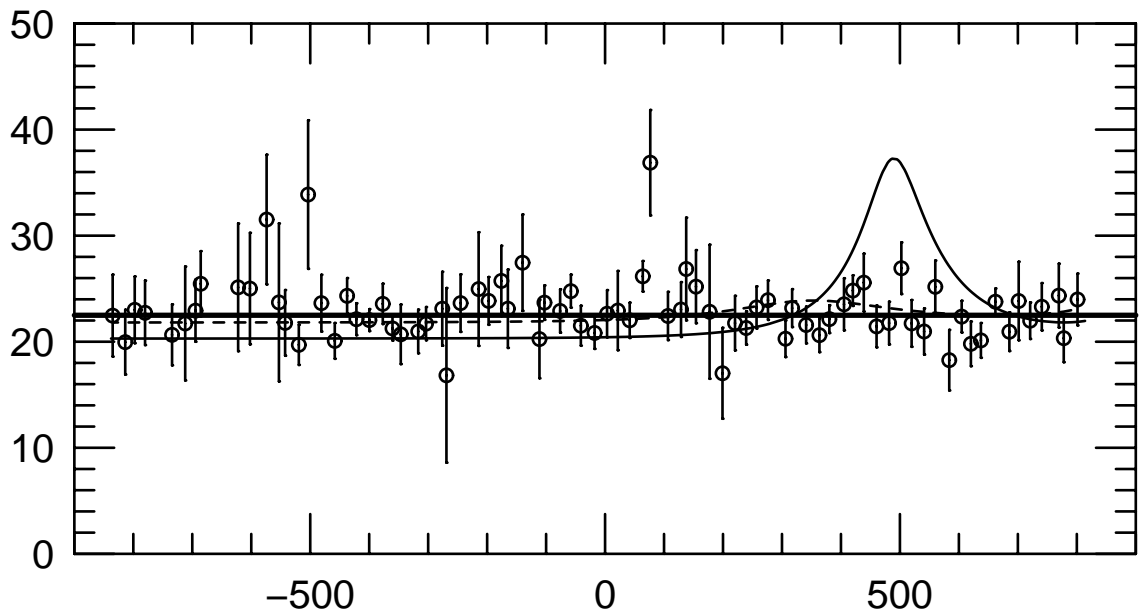
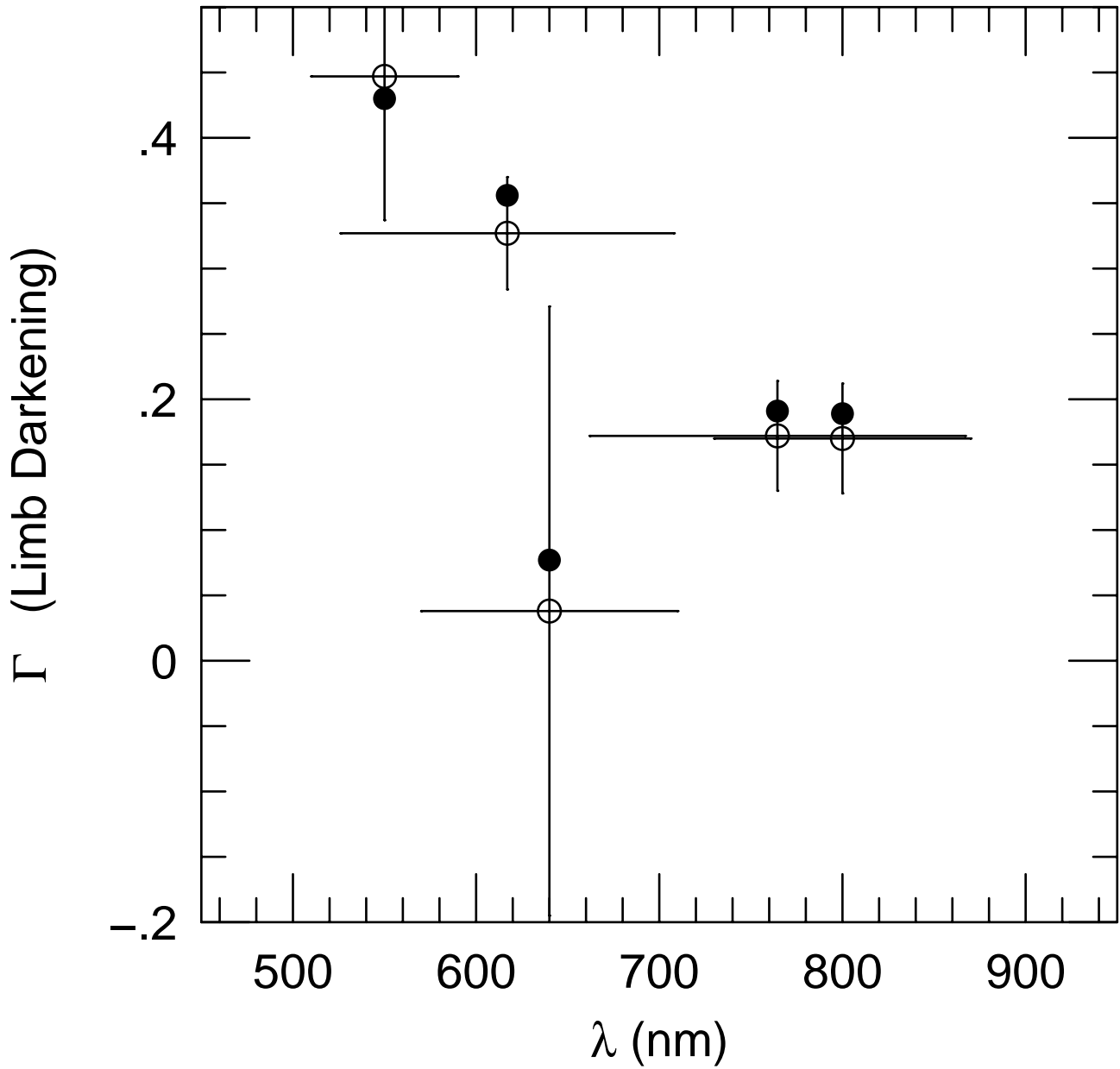
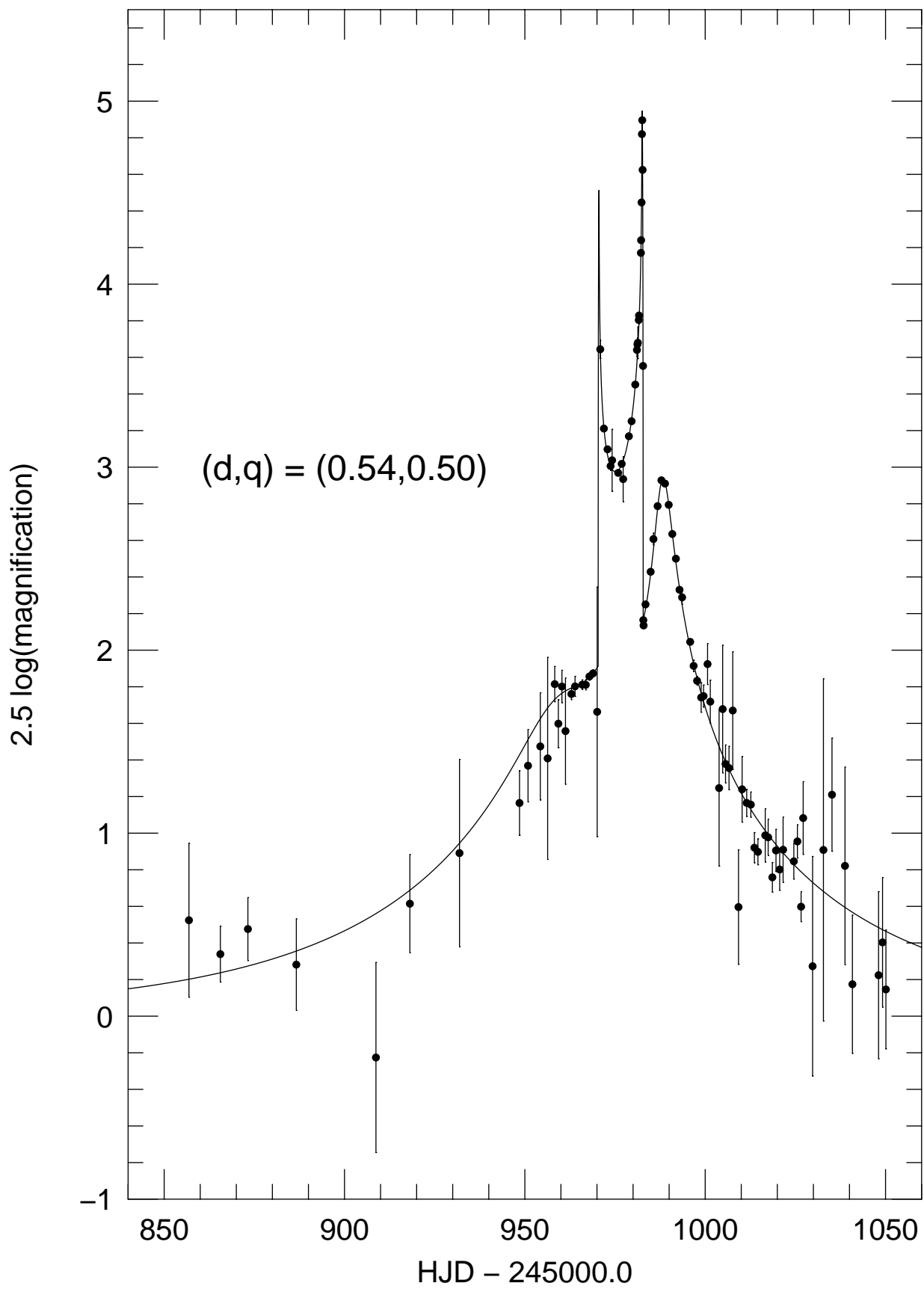
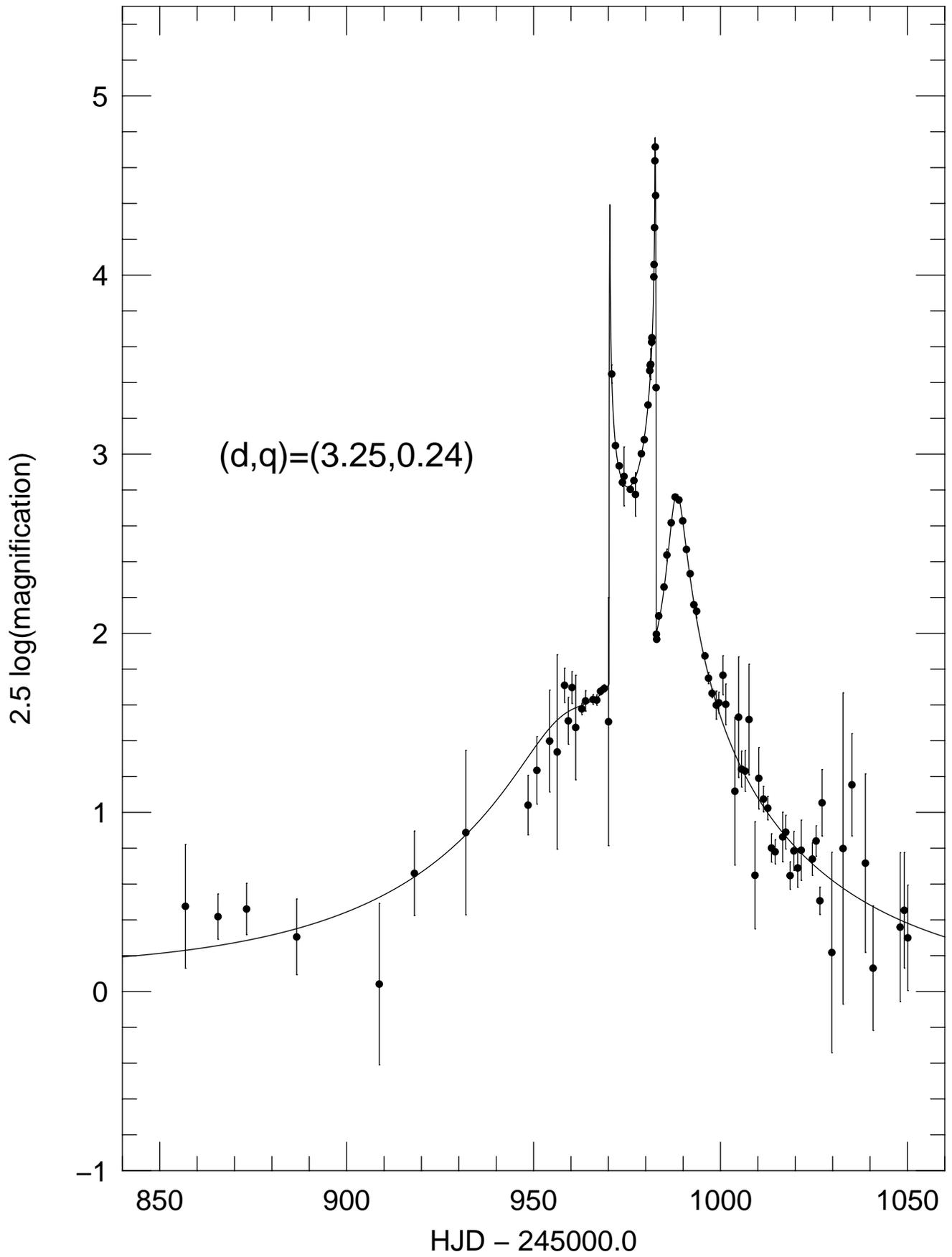
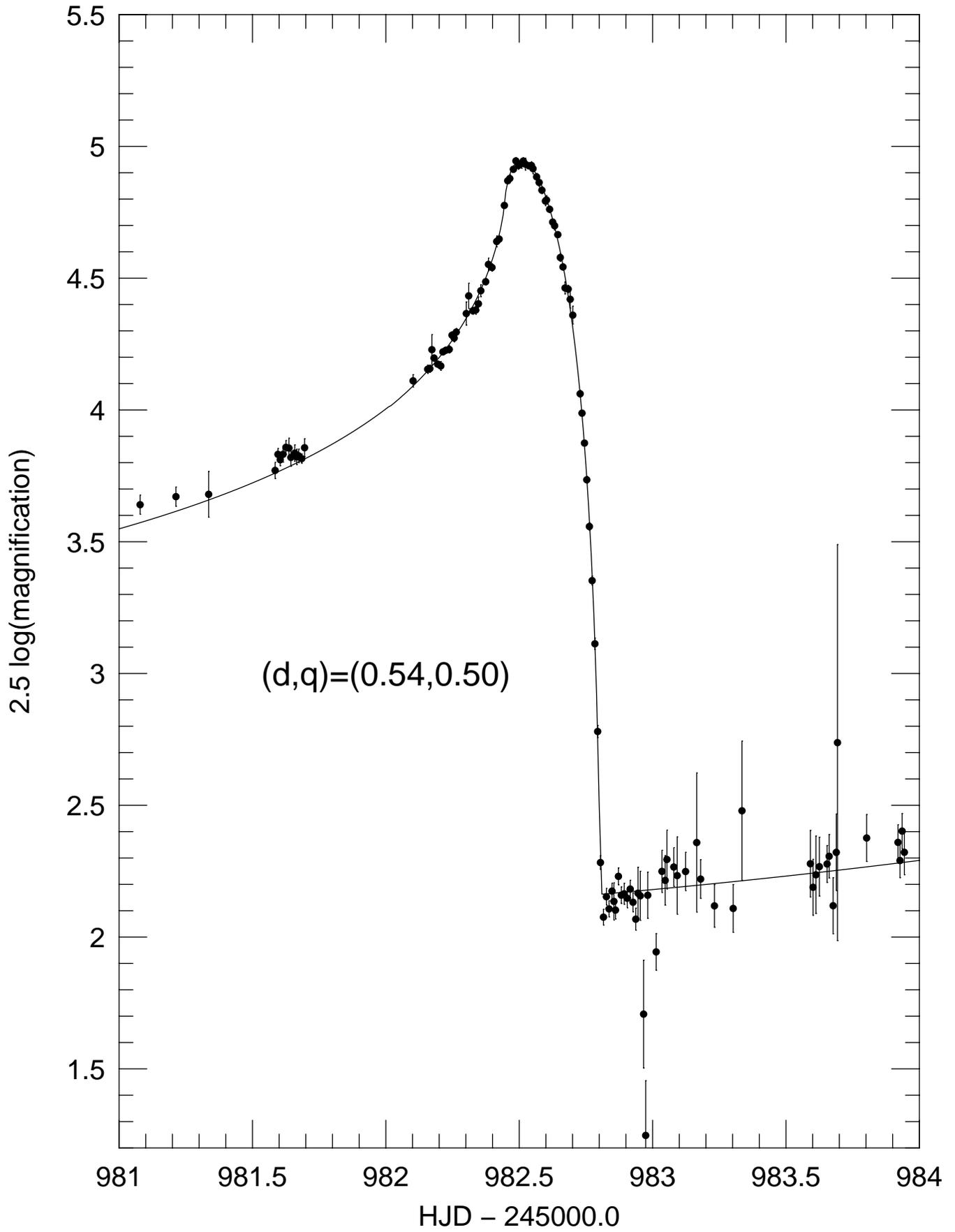
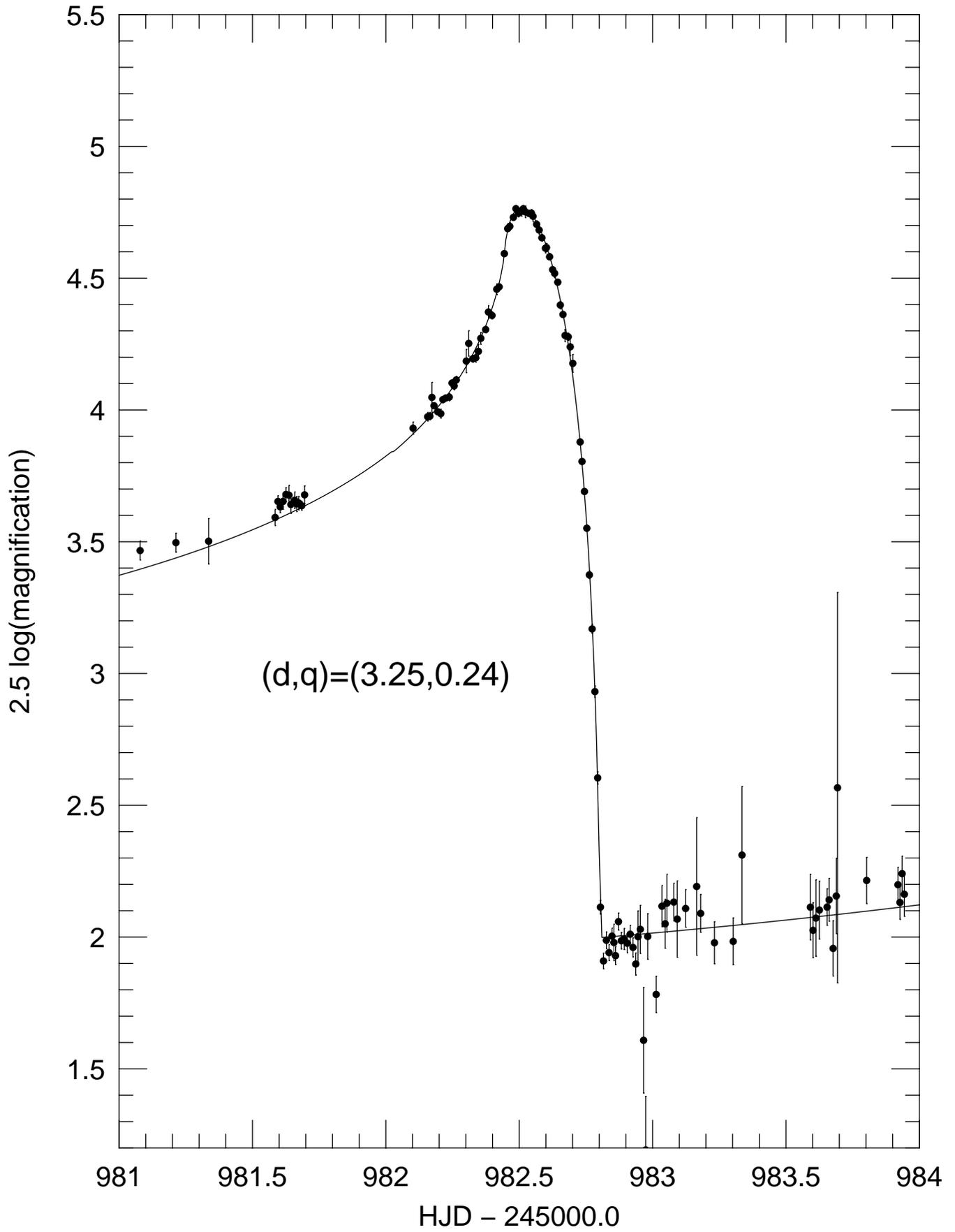
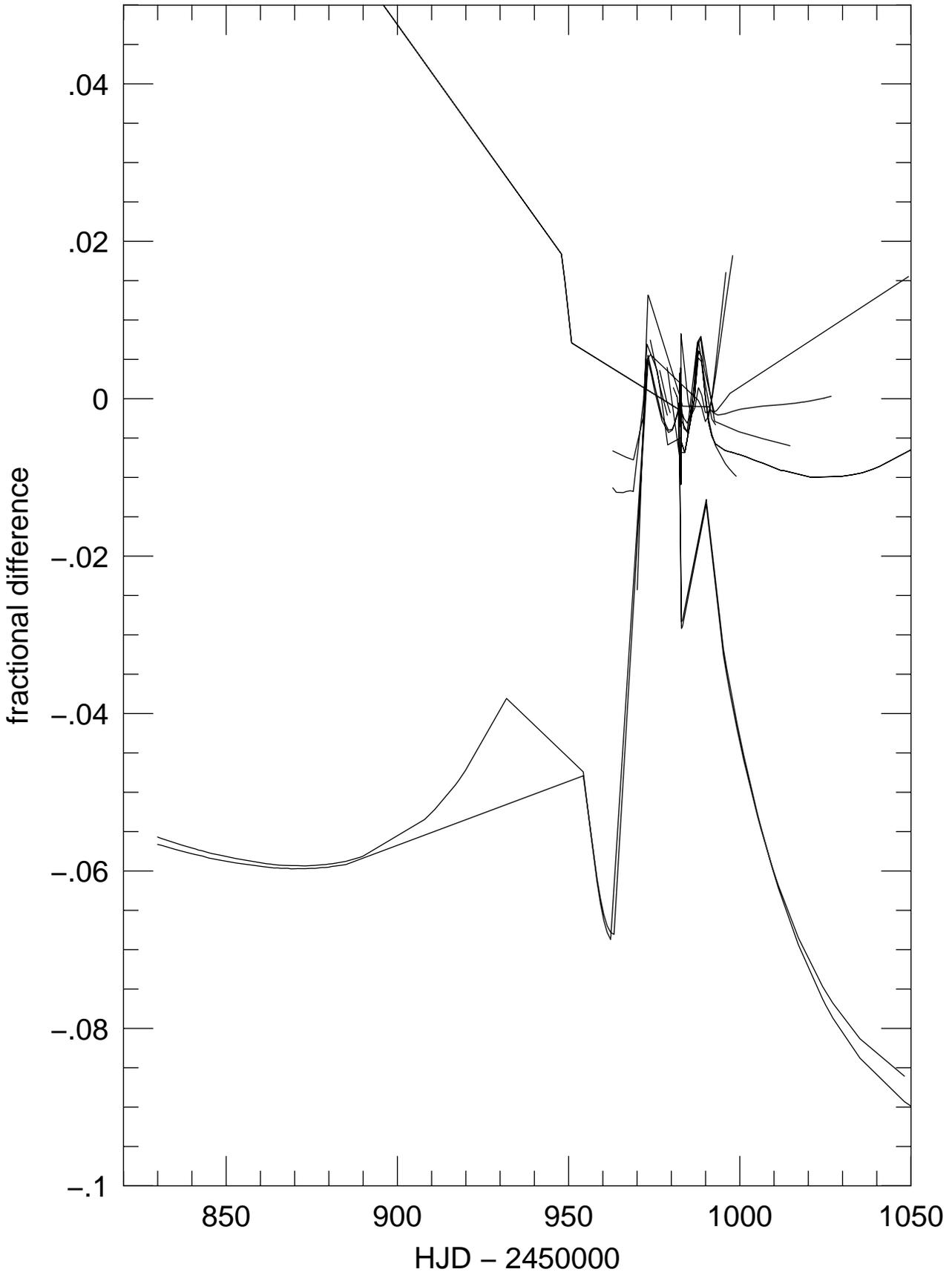
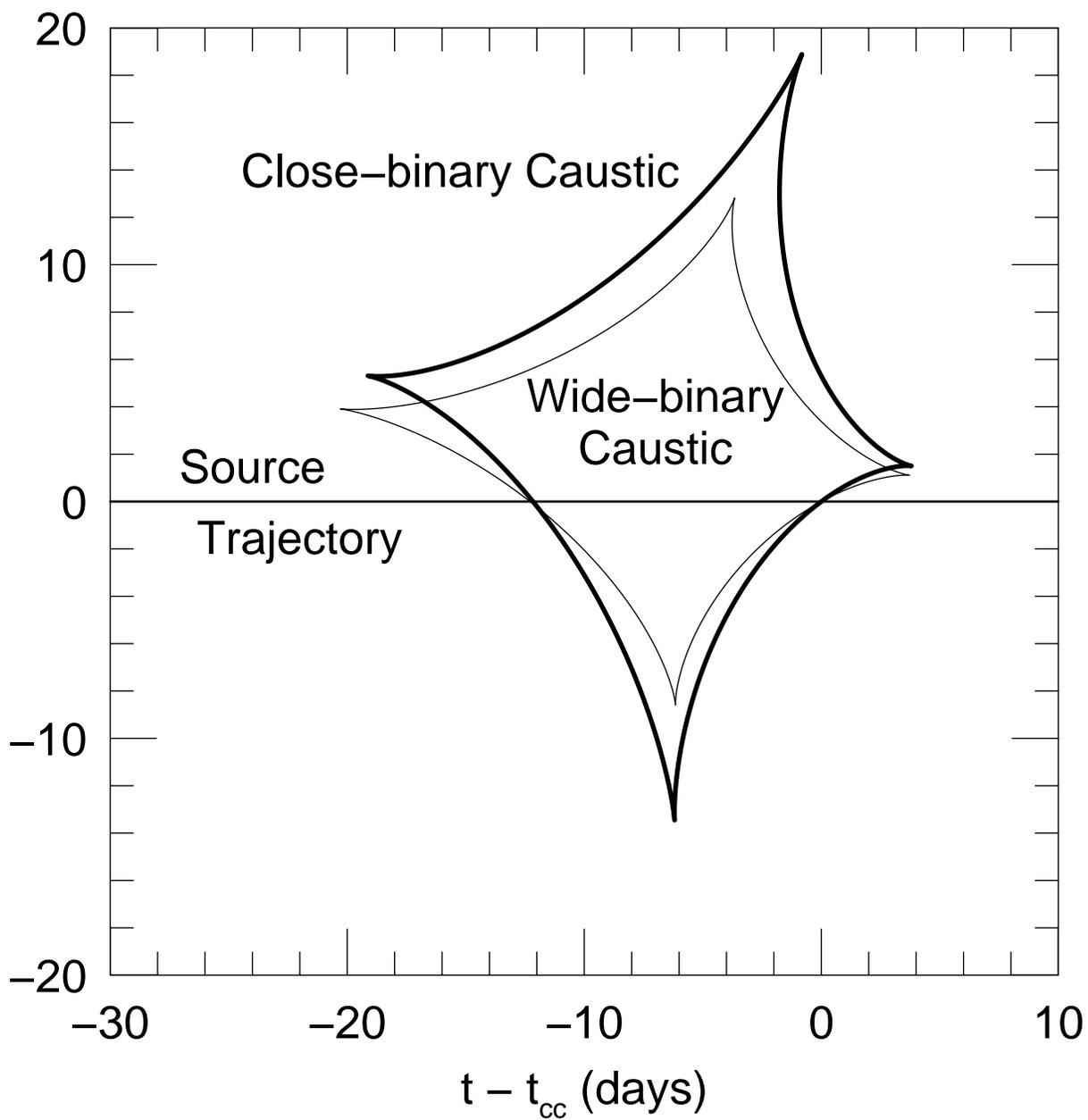
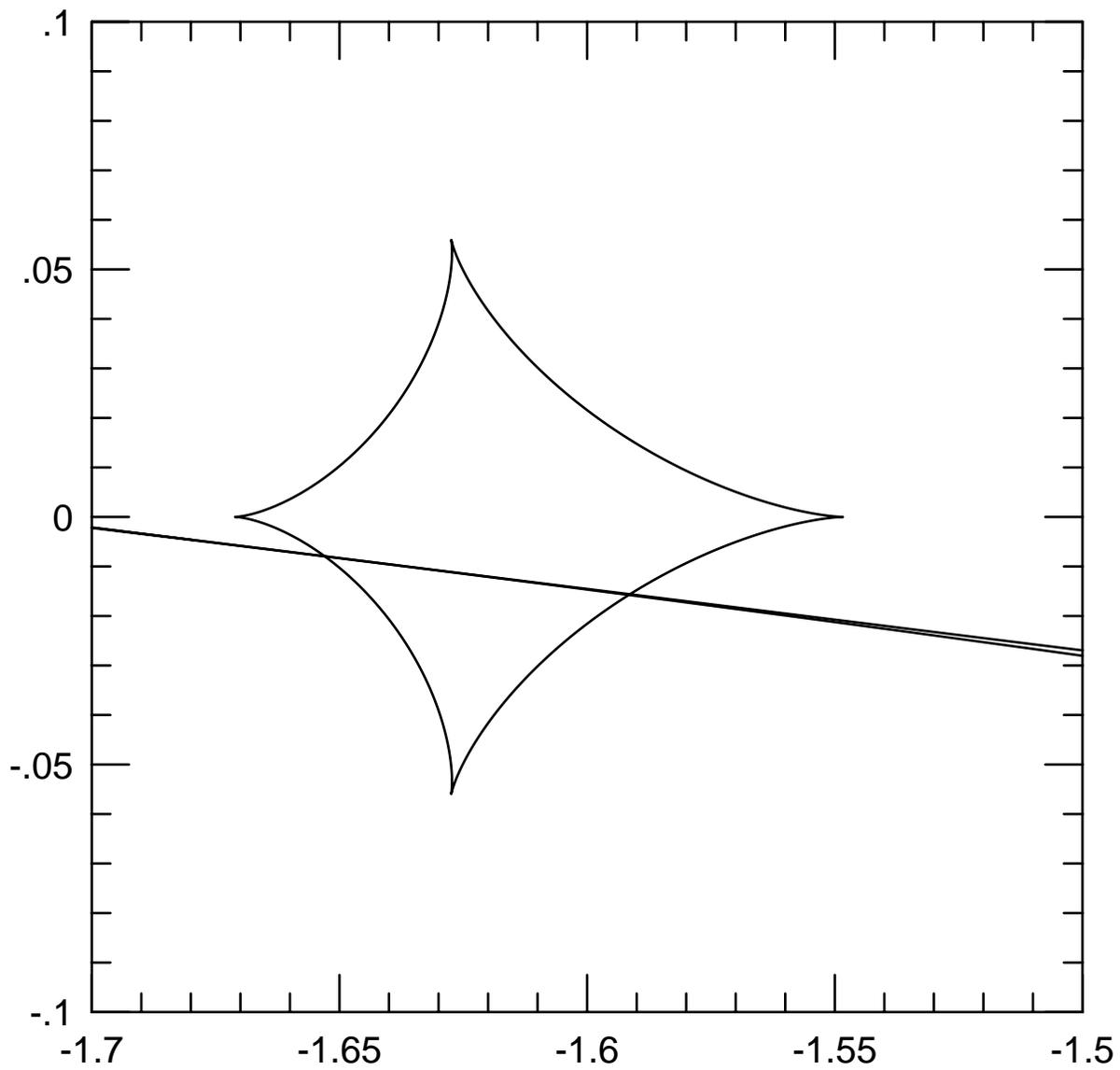
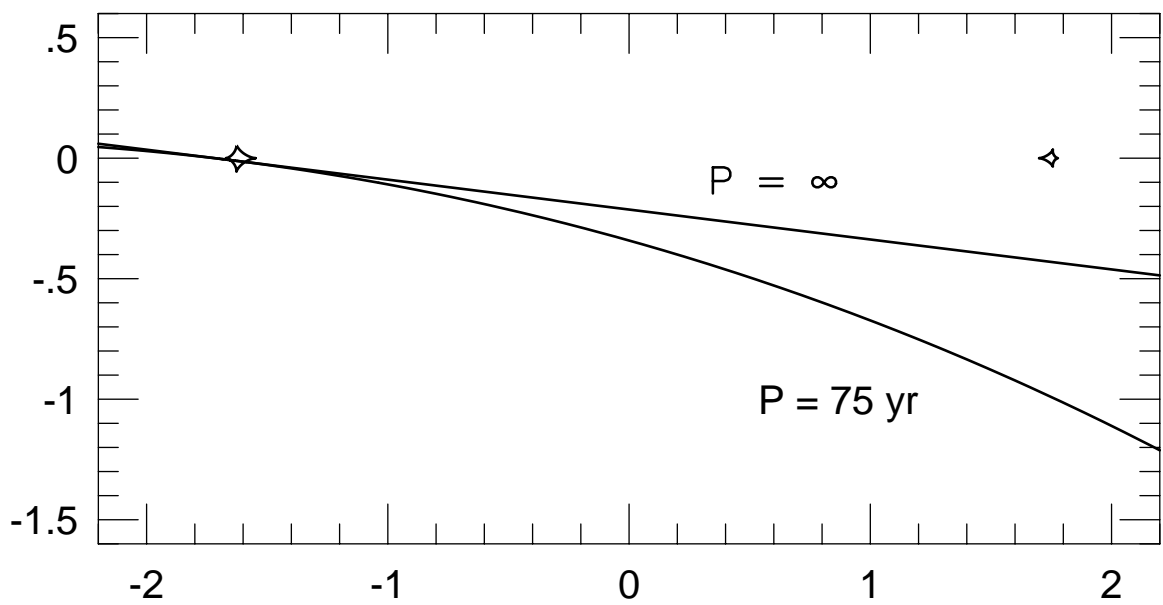
\end{document}